 \documentclass[12pt,preprint,usenatbib]{aastex}  %%%%%%%%%%
\def\dsm{$\mathrm{M}_\odot$}

\shorttitle{STUDY OF CMD OF NGC1651} \shortauthors{Li Z.M. et
al.}

\begin{document}


\title{EXPLANATION OF SPECIAL COLOR-MAGNITUDE DIAGRAM OF STAR CLUSTER NGC1651 FROM DIFFERENT MODELS}

\author{Zhongmu Li\altaffilmark{1,2}, Caiyan Mao\altaffilmark{1}, Li Chen\altaffilmark{1}}

\altaffiltext{1}{Institute for Astronomy and History of Science and
Technology, Dali University, Dali 671003, China;
zhongmu.li@gmail.com} \altaffiltext{2}{Key Laboratory for the
Structure and Evolution of Celestial Objects, Chinese Academy of
Sciences, Kunming 650011, China}

\begin{abstract}
The color-magnitude diagram (CMD) of
globular cluster NGC1651 has special structures including
a broad main sequence, an extended main sequence turn-off and an extended
red giant clump. The reason for such special CMDs remains unclear.
In order to test how different the
results from various stellar population assumptions are,
we study a high-quality CMD of NGC1651 from the Hubble
Space Telescope archive via eight kinds of models. Distance modulus,
extinction, age ranges, star formation mode, fraction of binaries,
and fraction of rotational stars are determined and then compared.
The results show that stellar populations both with and without age spread
can reproduce the special structure of the observed CMD. A
composite population with extended star formation from 1.8\,Gyrs ago to 1.4\,Gyrs ago,
which contains 50 per cent binaries and 70 per cent rotational stars, fits the
observed CMD best. Meanwhile, a 1.5\,Gyr-old simple population that consists of rotational stars can also fit the observed CMD well.
The results of CMD fitting are shown to depend strongly on
stellar population type (simple or composite), and fraction of rotators. If the member
stars of NGC1651 formed in a single star burst, the effect of
stellar rotation should be very important for the explanation of observed CMDs.
Otherwise, the effect may be small.
It is also possible that the special observed CMD is a result of the combined effects of stellar binarity, rotation and age spread.
Therefore, further work on stellar population
type and fraction of rotational stars of intermediate-age clusters are necessary
to understand their observed CMDs.
\end{abstract}


\section{Introduction}

Many intermediate-age star clusters, e.g., NGC1651,
show complicated color-magnitude
diagrams (CMDs), which include blue stragglers, a broad main sequence,
an extended main sequence turn-off (eMSTO) and an extended red clump(eRC). Such clusters
are found mainly in the Large Magellanic Cloud (LMC) \citep{mack07,
mack08, milo09,Piatti13},\cite{} and Small Magellanic Cloud (SMC)
\citep{girardi2009,Girardi13}. There has been a long history of the explanation
of such special CMDs. Factors considered as possible causes for CMDs with eMSTO and eRC features include spread of chemical abundance \citep{mack08,
goud09,piot05,piot07}, \cite{}capture of field stars \citep{mack08, goud09}, merger of
existing star clusters \citep{mack07}, formation of a second
generation of stars from the ejecta of first generation asymptotic
giant branch stars \citep{derc08,goud09}, binary stars (e.g.,
\citealt{milo09}), observational selection and uncertainty effects
\citep{kell11}, mixture of stars with and without overshooting
\citep{Girardi11}, differential reddening \citep{platais12}, age
spread \citep{Girardi11,Richer13}, stellar rotation \citep{bast09},
and combination of binaries and rotation \citep{Li12}. Meanwhile,
some other studies, including \cite{mucc08}, \cite{Goudfrooij11b},
\cite{Goudfrooij11a}, \cite{glat08}, \cite{rube10},
\cite{Girardi11}, and \cite{Girardi13}, pointed out some challenges to
these assumptions. Finally, a spreading in age (e.g.,
\citealt{Girardi11}) and stellar rotation (e.g.,
\citealt{bast09,Li12}) are thought to be the most probable causes for
special CMDs with eMSTO and eRC, because only such models can
reproduce the eMSTO part sufficiently well.

Some special CMDs of clusters have been compared with
theoretical models, but no work compares the results
from a few different stellar population models simultaneously.
Such an approach can, however, give new
insight into the relative importance of various factors involved in the
observed CMD. This work aims to supply such an attempt via a typical cluster,
NGC1651. This is a red globular cluster in the LMC, which is
situated 3 kpc southwest of the Bar in a region that appears fairly
free of recent star formation. The CMD of NGC1651 has been studied before, e.g.,
\cite{Mould1986}, \cite{Mould1997}, \cite{Brocato2001}, \cite{Sarajedini2002}, and \cite{groc07}.
Although not very clear, special CMD structures with eMSTO and eRC,  which have first been
described by \cite{mack07} and \cite{girardi2009}, can be seen in the results of \cite{Brocato2001}.
The special CMD of NGC1651 was recently confirmed by the present authors using data from the Hubble Space Telescope (HST) archive,
in which blue stragglers, broad MS, eMSTO and eRC structures seem very clear.
We intend to study this CMD via various model assumptions. The study will help
to clarify the roles of various factors in the explanation of observed CMDs
and in the determination of cluster parameters.
In the most complex model, the effects of binaries, age spread,
rotating stars, and star formation history are taken into account,
besides those of distance, extinction, and metallicity.
The structure of this paper is as
follows: Sections 2 and 3 introduce the observed and theoretical
CMDs; Section 4 introduces the technique of CMD fitting;
Section 5 presents the best-fit results from different models;
and finally, Section 6 contains a summary and discussion.

\section{Observed CMD of NGC1651}
\subsection{Data and photometry}
The data of NGC1651 are obtained from the HST
archive, and consist of total exposures of 500 seconds in the F555W
and F814W filters of Wide Field Planetary Camera 2 (WFPC2).  In
order to guarantee the data quality, we handle the data using a
stellar photometry package (HSTphot, \citealt{Dolphin2000})
specially designed for use with HST WFPC2 images. The magnitudes are
automatically transformed to standard $UBVRI$ magnitudes by HSTphot.
We are shown that stars brighter than 19.0\,mag are not measured with the long exposures.
Although the stars distribute in wide color and magnitude ranges, only those near the turn-off are taken to serve the purpose of this paper.
We finally obtain a CMD that consists of 4244 stars with $V$
magnitude between 18.5 and 25.5\,mag, and $(V-I)$ color between 0
and 1.8\,mag. The number of stars is larger than previous works
(e.g., \citealt{Mould1997,Sarajedini2002}), and the CMD seems
clear, which is helpful for our detailed study.
The CMD part fainter than 25.5\,mag is not used because of its high incompleteness ($>$ 40 \%).

\subsection{Photometric errors and completeness}
A kind of photometric uncertainty has been reported
by HSTphot, but it is obviously less than the real error
because simulated CMDs are much narrower than the observed
one when the error reported by HSTphot is taken into account.
Furthermore, the sample seems incomplete, because the stars of
NGC1651 are in a crowded field. This will surely lead to some extra errors.
We therefore perform artificial
star tests (ASTs) to characterize the photometric errors and
completeness of stars. Via this technique,
most uncertainties caused by crowding and the photometry process can be estimated.
The ASTs are performed as follows.
First, a large number of images that include more than 10$^{\rm 5}$ artificial stars in the
same color and magnitude ranges as the observed ones are generated.
For convenience, the real magnitudes of artificial stars are recorded as input magnitudes.
Then the generated images are processed
by HSTphot to find out stars and measure their magnitudes.
The measured magnitudes are recorded as recovered magnitudes.
Finally, the input and recovered magnitudes of artificial stars are compared, to
characterize the photometric errors, and the input and recovered
star numbers are compared, to give star completeness.
Fig. 1 shows the photometric errors as a function of $V$ and $I$
input magnitudes, and Fig. 2 shows the completeness of different CMD
parts of NGC1651. We see that photometric errors mainly depend on
recovered magnitudes. The fainter the magnitudes, the larger the spread of photometric errors.
In addition, we find that the completeness of various CMD parts are
different. Fainter areas have lower completeness on average.
Thus we applied position-dependent completeness to the
observed CMD. Note that the results for where observed stars are located are highlighted in Fig. 2,
in order to help readers understand the observed CMD better.
Fig. 3 shows some compensating stars that are randomly added into the observed CMD according to star completeness.
Because added stars are far fewer than the observed stars, they do not affect the final results too much.
The final observed CMD of NGC1651, which has included compensating stars, is shown by Fig. 4.
This CMD consists of 4803 stars and is much clearer than
what is used by previous works \citep{Mould1997,Sarajedini2002}.
We can see broad MS, eMSTO and eRC (see Sect. 5 for details) structures clearly from this CMD. It is therefore a good example for our detailed study.

\begin{figure} %Fig 1.
\centering
\includegraphics[angle=-90,width=0.48\textwidth]{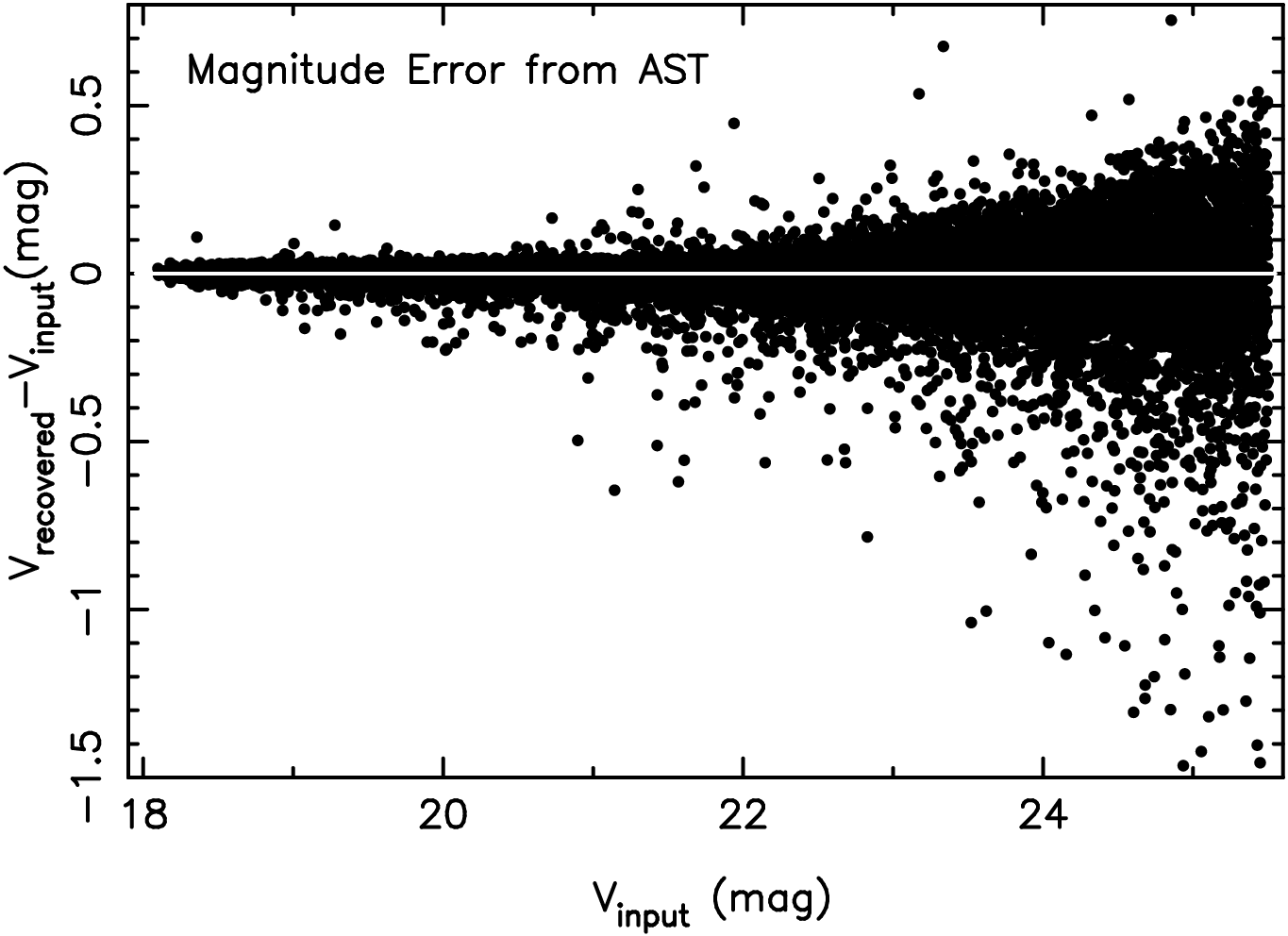}
\includegraphics[angle=-90,width=0.48\textwidth]{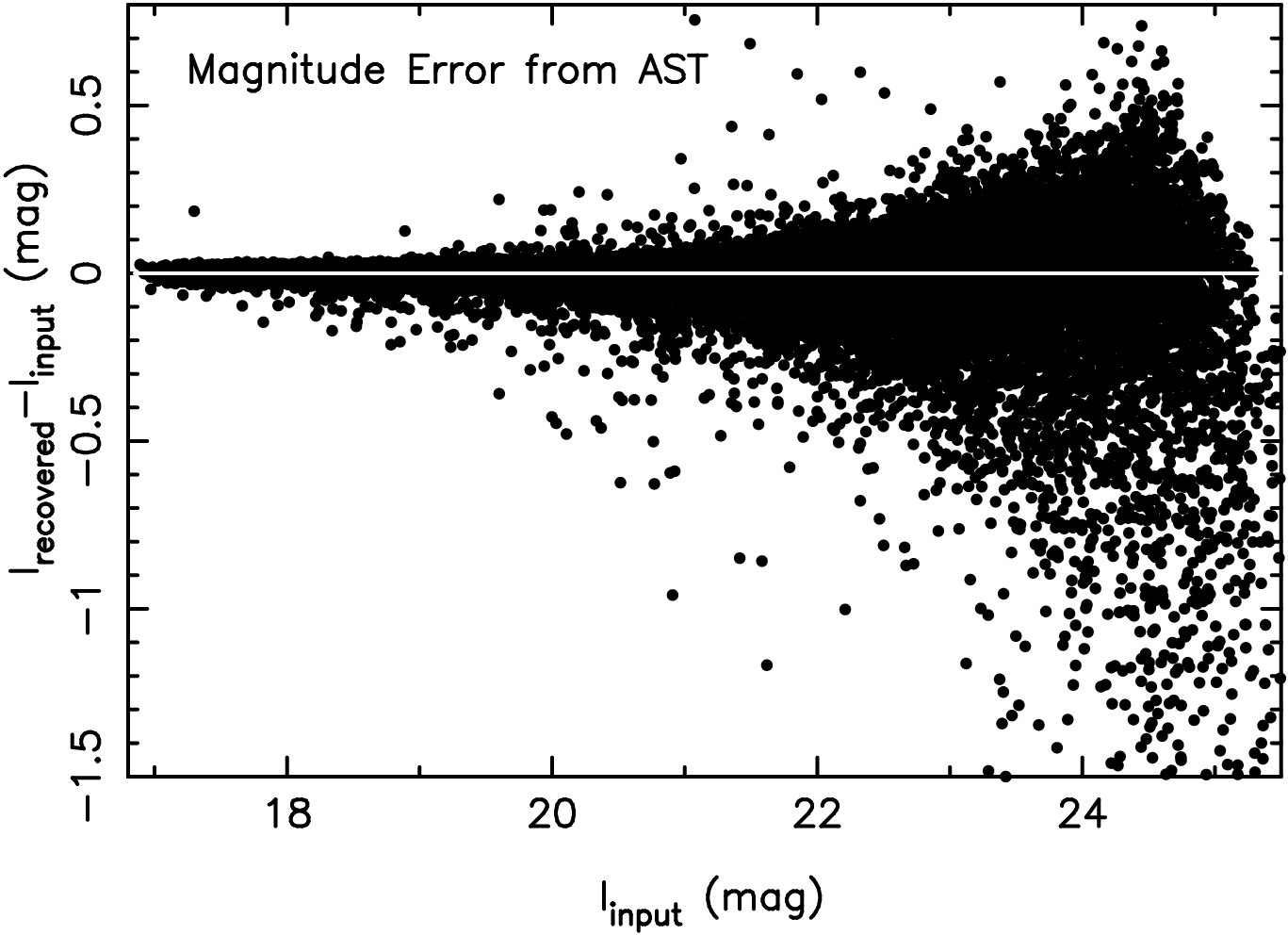}
\caption{Magnitude error as a function of measured magnitude of cluster NGC1651.
Errors are estimated using the ASTs. Left and right panels are for $V$ and $I$ magnitudes, respectively.}
\end{figure}

\begin{figure} %Fig 2.
\centering
\includegraphics[angle=-90,width=0.9\textwidth]{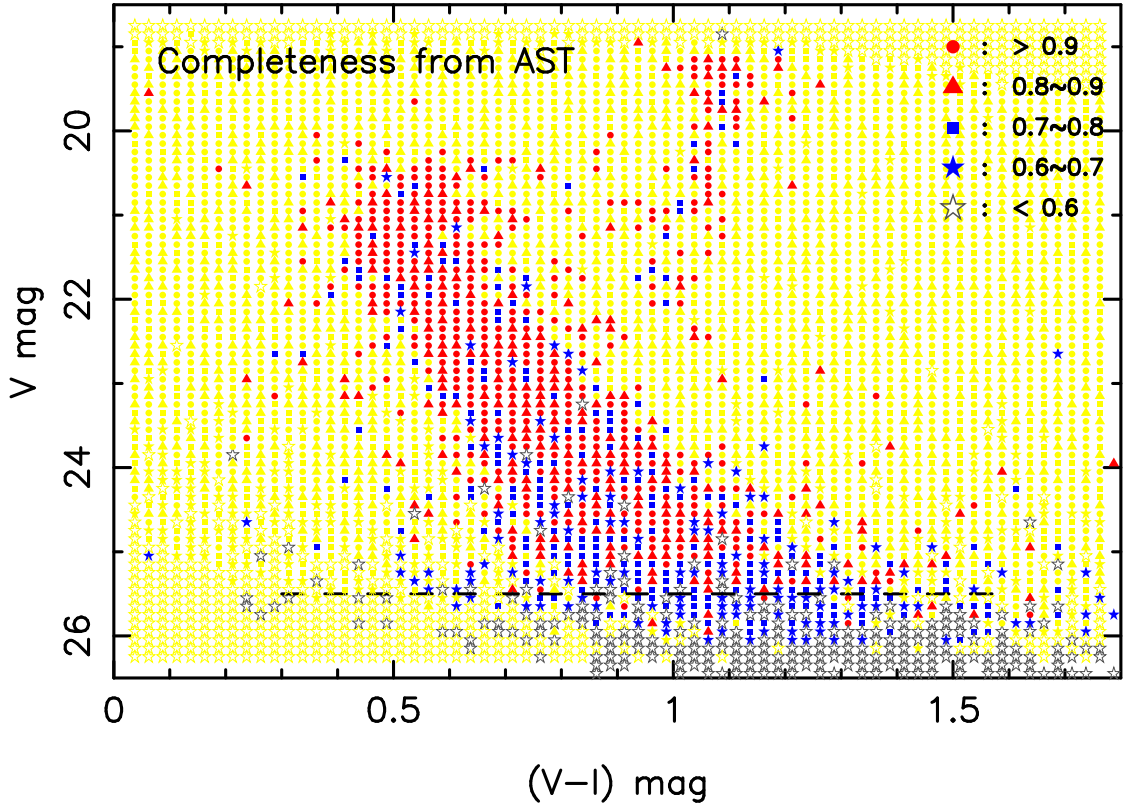}
\caption{Star completeness of NGC1651. Only stars over the dashed line (25.5\,mag) are used for our study,
because most such grids have completeness greater than 60 per cent. The results for where observed stars are located are highlighted by red, blue and black colors,
and that for the other part is plotted in yellow.}
\end{figure}

\begin{figure} %Fig 3.
\centering
\includegraphics[angle=-90,width=0.9\textwidth]{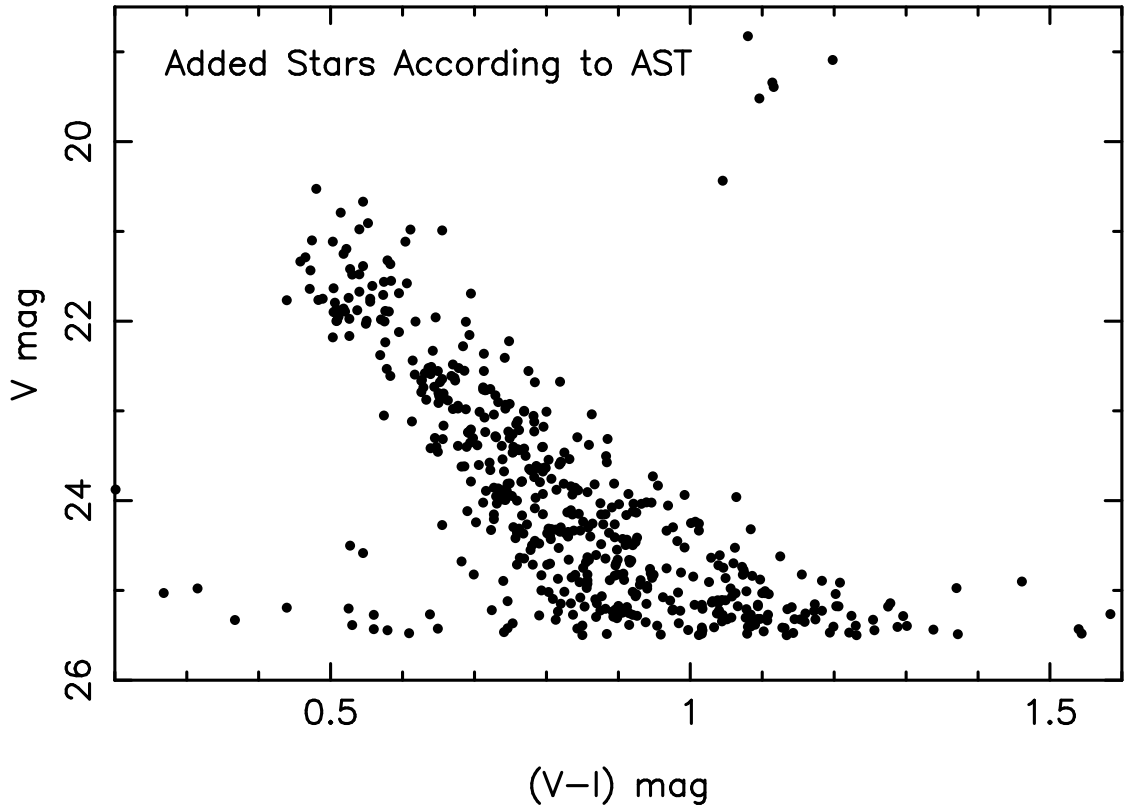}
\caption{Stars randomly added into observed CMD according to AST completeness.
Stars fainter than 25.5\,mag are not shown because they are not used in this work.}
\end{figure}

\begin{figure} %Fig 4.
\centering
\includegraphics[angle=-90,width=0.9\textwidth]{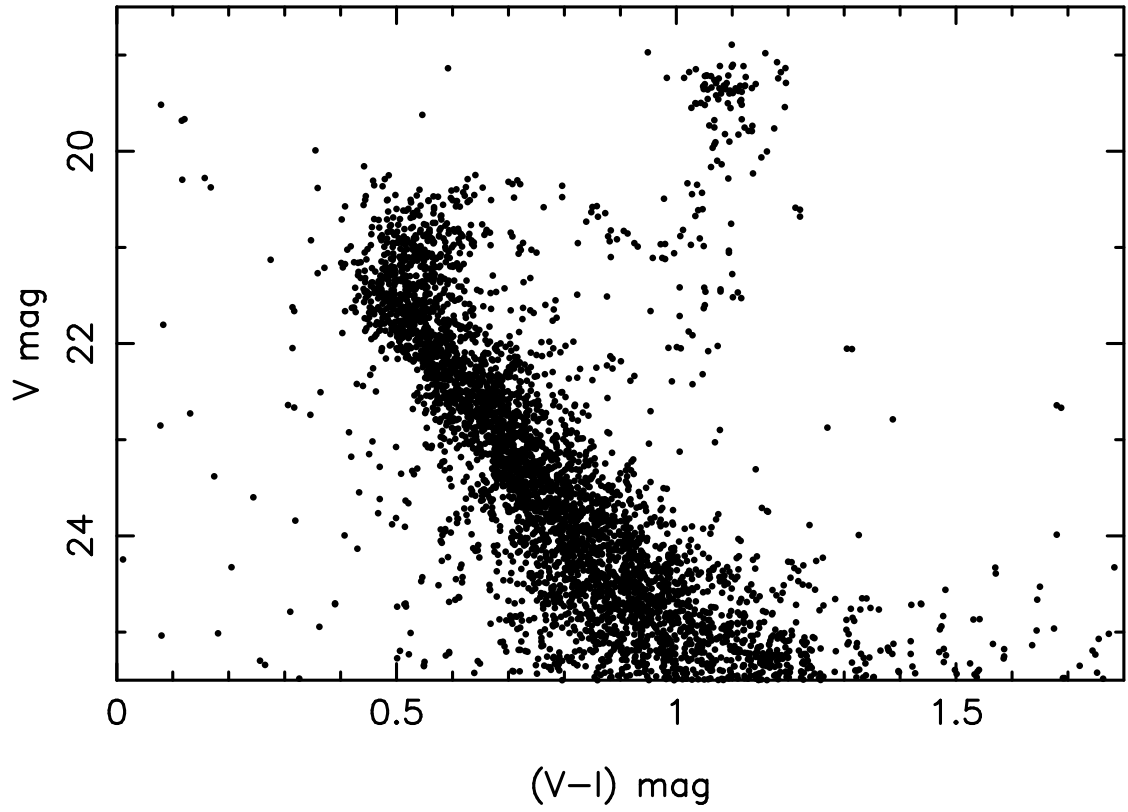}
\caption{Observed CMD of NGC1651.}
\end{figure}

\section{Synthetic CMDs}
We construct synthetic CMDs following previously cited works (e.g.,
\citealt{zhang04,Li2008mn,Li2008apj,Li2012}). Simple and composite stellar populations
(SSPs and CSPs) are built by taking the initial mass function (IMF)
of \cite{Salpeter1955} with lower and upper mass limits of 0.1 and
100 \dsm{}, respectively. We build both single star stellar
populations (ssSPs) and binary star stellar populations (bsSPs). The
difference between the two kinds of populations is that bsSPs contain
some binaries but ssSPs do not. In each binary, the mass of the
primary component is generated following the selected IMF, and the
mass of the secondary component is calculated by taking a random
secondary-to-primary mass ratio ($q$), which obeys a uniform
distribution within 0--1. The eccentricity ($e$) of each binary is
given randomly within 0-1. The separations ($a$) of two components
are given by:
\begin{equation}
an(a)=\left\{
 \begin{array}{lc}
 \alpha_{\rm sep}(a/a_{\rm 0})^{\rm m} & a\leq a_{\rm 0};\\
\alpha_{\rm sep}, & a_{\rm 0}<a<a_{\rm 1},\\
\end{array}\right.
\end{equation}
where $\alpha_{\rm sep}\approx0.070$, $a_{\rm 0}=10R_{\odot}$,
$a_{\rm 1}=5.75\times 10^{\rm 6}R_{\odot}=0.13{\rm pc}$ and
$m\approx1.2$ \citep{han95}. This leads to about 50 per cent stars
in binaries with orbital periods less than 100 yr. Although the
typical binary fraction of LMC clusters is about 35 per cent
\citep{elson98,bast09}, the values for various clusters may be
different. Thus some population models with different binary
fractions are built in this work. We remove some random binaries to
make the binary fraction equal to what we need.

After the generation of the star sample, we calculate the evolution of
stars using the rapid stellar evolution code of \cite{Hurley98} and \cite{Hurley02}
(hereafter Hurley code), which can evolve both single stars and binaries. In
binary evolution, most binary interactions such as mass transfer,
mass accretion, common-envelope evolution, collisions, supernova
kicks, angular momentum loss mechanism, and tidal interactions are
taken into account. The typical uncertainty of this code is about 5
per cent in stellar luminosity, radius and core mass, which affects the results slightly.

Because Hurley code does not take stellar rotation
into account, we add the effect of rotation on effective
temperature, luminosity and MS lifetime to massive ($>$ 1.7
\dsm{}) stars, using a recent result of evolution of rotating stars \citep{georgy2013}.
The database of \cite{georgy2013} is particularly
useful for constructing synthetic populations of stars, accounting
for mass, rotation, and metallicity distributions. It includes an accurate computation
of the angular-momentum and stellar-wind anisotropy.
The changes of surface temperature and luminosity, which are caused by stellar rotation, are calculated by
comparing the evolutionary tracks of rotating stars to those of non-rotating ones.
The changes can be described as a function of stellar metallicity, mass, and initial rotation rate.
Besides changing the evolutionary track, rotation significantly affects the MS lifetime of stars.
Because of centrifugal forces, rotating stars behave like non-rotating stars of lower mass (see also \citealt{ekstrom2008,meynet2000}).
We therefore calculate the MS lifetime change of rotating stars using the result of \citet{georgy2013}.
Stars with rotation rate (the ratio of the actual angular velocity to the critical one, $\omega$, see \citealt{georgy2013}) larger than 0.3 usually have a lifetime change greater than 15\%,
and the most rapid rotators may reach values of 30 -- 35\%.
Because of the lack of evolutionary tracks of stars with $Z$=0.008, stellar metallicity is interpolated to
the typical value of LMC clusters to fit the needs of this work. In addition,
some random values are assigned to the stars of a population following the observed results of \citet{royer07} (similarly, \citealt{zorec2012}), because the rotation rate of stars has obvious distributions. Finally, the stellar evolutionary parameters ([Fe/H],
T$_{eff}$, $\log g$, $\log L$) are transformed into colors and
magnitudes using the atmosphere library of \cite{leje98}.
The photometric errors derived from ASTs are applied to the simulated stars when comparing the synthetic CMDs with the observed one.

\section{CMD fitting and application to simple CMDs}
In this work we want to determine the distance modulus, color excess, binary fraction,
rotational star fraction, and star formation mode of clusters.
In order to determine these effectively,
we apply a new CMD fitting method.
The new technique searches for best-fit parameters by
comparing the star fraction in every part (or grid) of a CMD.
An observed CMD is divided into many grids via selected color and magnitude intervals.
The goodness of fit is judged by the average difference of grids that contain observed or theoretical stars (hereafter effective grids),
when taking the weight of each grid into account.
For convenience, we defined a parameter, weighted average difference ($WAD$), to denote the goodness of fit.
It can be calculated by
\begin{equation}\label{eq3}
    WAD = \frac{\Sigma [{\omega_i.|f_{ob}-f_{th}|}]}{\sum\omega_i}~,
\end{equation}
where $\omega_i$ is the weight of $i$th grid, and
$f_{ob}$ and $f_{th}$ are star fractions of observed and theoretical
CMDs in the same grid. $\omega_i$ is calculated as
\begin{equation}\label{eq4}
    \omega_i = \frac{1}{|1-C_i|} = \frac{1}{\sigma_i}~,
\end{equation}
where $C_i$ is the completeness of $i$th grid that is given by AST, and $\sigma_i$ means star fraction uncertainty.

Although various indicators are used for the goodness of CMD fitting by different works,
there are obvious advantages to take $WAD$ in this work, which counts stars but is different from  the approach used by \cite{bert03}.
Firstly, $WAD$ can give us a rough estimation of the difference between observed and synthetic CMDs.
In other words, we believe that $WAD$ has more physical meaning than others, such as the widely used $\chi^2$,
which is suitable for Gaussian observational error distributions, and Poisson equivalent $\chi^2$ (hereafter $\chi_e^2$),
which is suitable for Poisson error distributions \citep{Dolphin2002}.
Secondly, $WAD$ has considered most information about the observed CMD, including  distributions of magnitude errors,
completeness of stars, and star fraction difference of observed and theoretical CMDs.
Thirdly, unlike the $\chi_e^2$ method of \cite{Dolphin2002}, which requires there to be observed star in very grid (in the calculation of $\chi_e^2$, i.e., Equation 5 in \cite{Dolphin2002}, $log(f_{th}/f_{ob})$ is meaningless when $f_{ob}$ = 0), our method, allows grids to have an absence of observed stars when calculating the goodness indicator.
In order to help readers to understand the relation between $WAD$ and other indicators,
Fig. 5 compares $WAD$ with two other indicators ($\chi^2$ and $\chi_e^2$).
Note that the data are taken from the CMD fitting of NGC1651 using a few different models.
We see that $WAD$ increases with $\chi^2$ as a whole, which suggests that they may give similar results.
However, $WAD$ and $\chi_e^2$ will possibly give different results, because there is no correlation between them.
Note that $\chi^2$ is calculated by
\begin{equation}\label{eq4}
    \chi^2 = \Sigma \frac{(f_{ob}-f_{th})^2}{(1-C)^2},
\end{equation}
and $\chi_e^2$ by
\begin{equation}\label{eq4}
    \chi_e^2 = 2\Sigma [(f_{ob}-f_{th})+f_{th}.log(f_{th}/f_{ob})].
\end{equation}

\begin{figure} %Fig 5.
\centering
\includegraphics[angle=-90,width=0.8\textwidth]{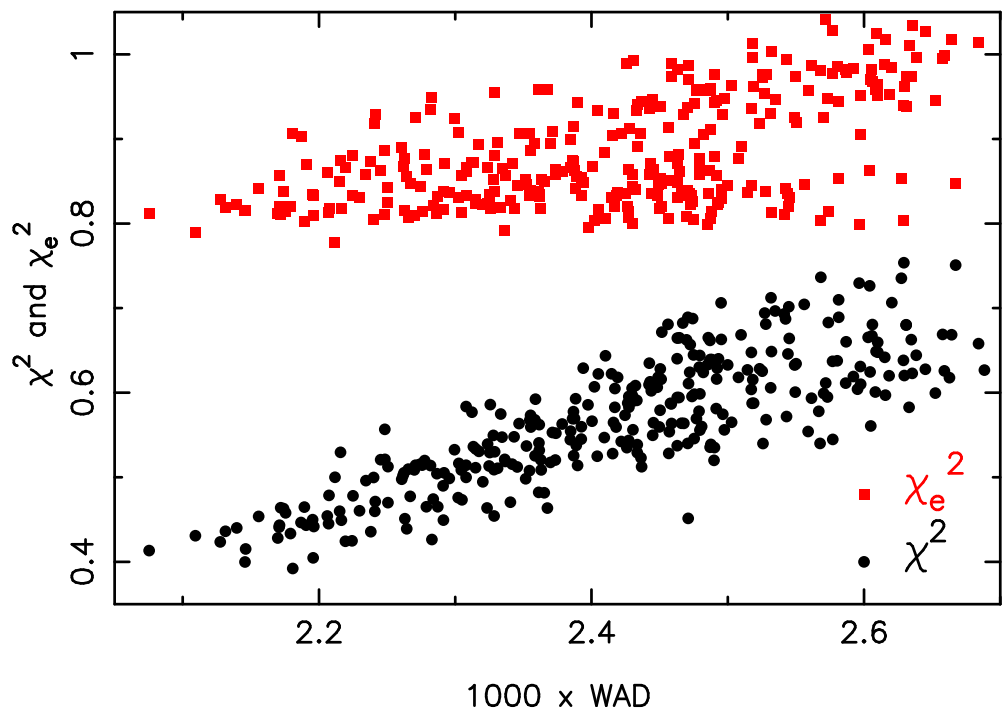}
\caption{Comparison of $WAD$ to two widely used goodness indicators of fit, $\chi^2$ and $\chi_e^2$.
The data are taken from some test runs of CMD fit of cluster NGC1651. Black points and red squares are for $\chi^2$ and $\chi_e^2$, respectively.}
\end{figure}

When we test the above three kinds of fitting methods using a few artificial clusters
($Z$ = 0.008; age = 0.9, 1.2, 1.5, 1.8, 2.1\,Gyrs; $f_{\rm b}$ = 0.0, 0.3, 0.6),
it is shown that $WAD$ and $\chi^2$ methods usually give correct results, but $\chi_e^2$ method
does not report correct results for some cases.
In particular, $WAD$ fitting can determine the stellar ages of artificial clusters accurately,
although there is an uncertainty of about 0.1 in $f_{\rm b}$.
When we apply $WAD$ fitting to two simple clusters (NGC1261 and NGC2257),
which do not have an eMSTO structure,
the best-fit models agree well with observed CMDs as a whole (Fig. 8).
We see that a few stars, in particular those on the horizontal branch (HB), are not fitted well.
This may result from the still large uncertainties in the late phases of stellar evolution.
In deed, HBs are very tricky to model, because of mass loss.
The test therefore suggests that the $WAD$ method is reliable for CMD studies.
Note that the data of two clusters are also taken from HST,
and one can check their magnitude versus error relations via Figs. 6 and 7.
In fact, magnitude errors affect the shapes of CMDs visibly.

\begin{figure} %Fig 6.
\centering
\includegraphics[angle=-90,width=0.45\textwidth]{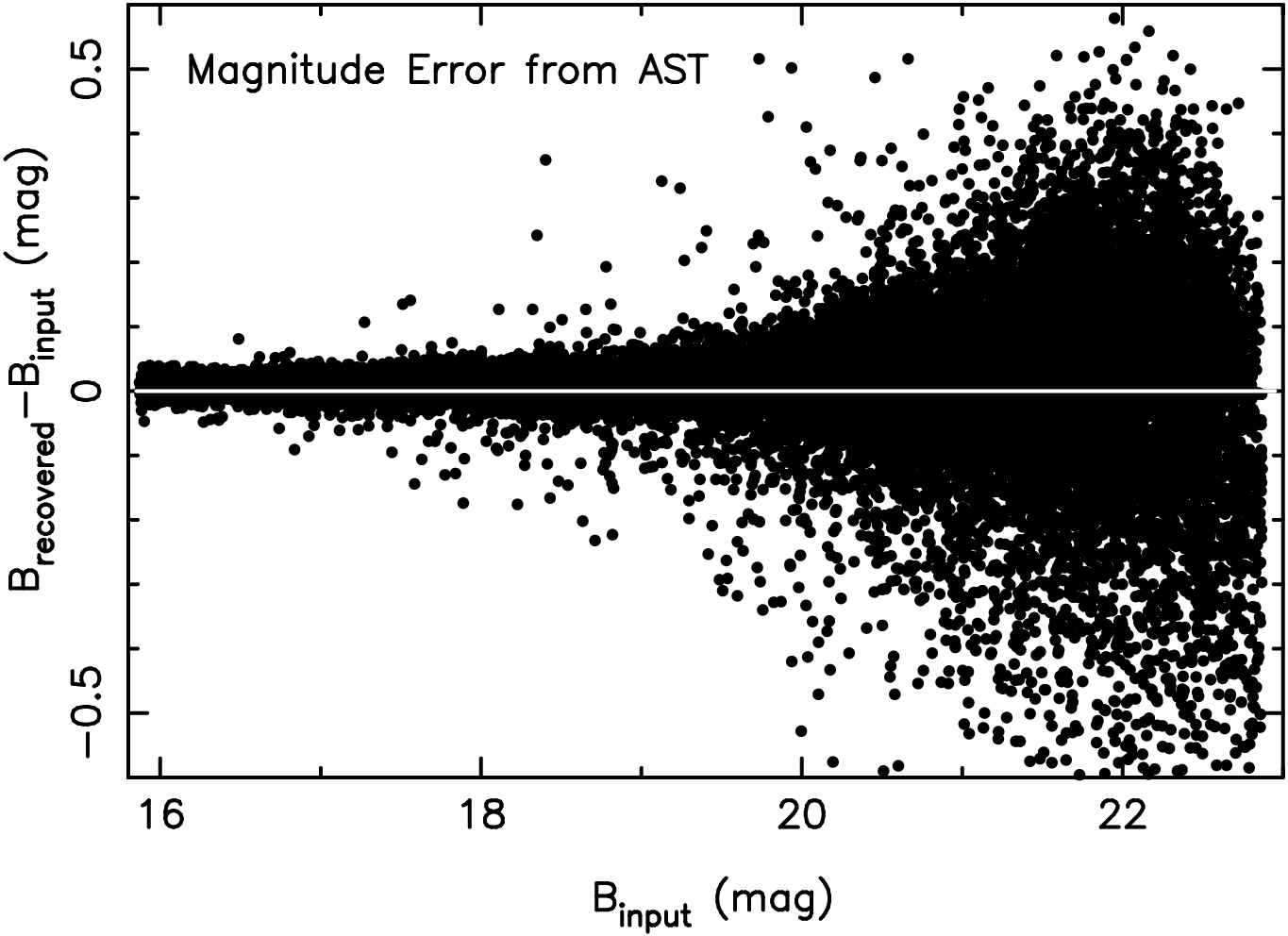}
\includegraphics[angle=-90,width=0.45\textwidth]{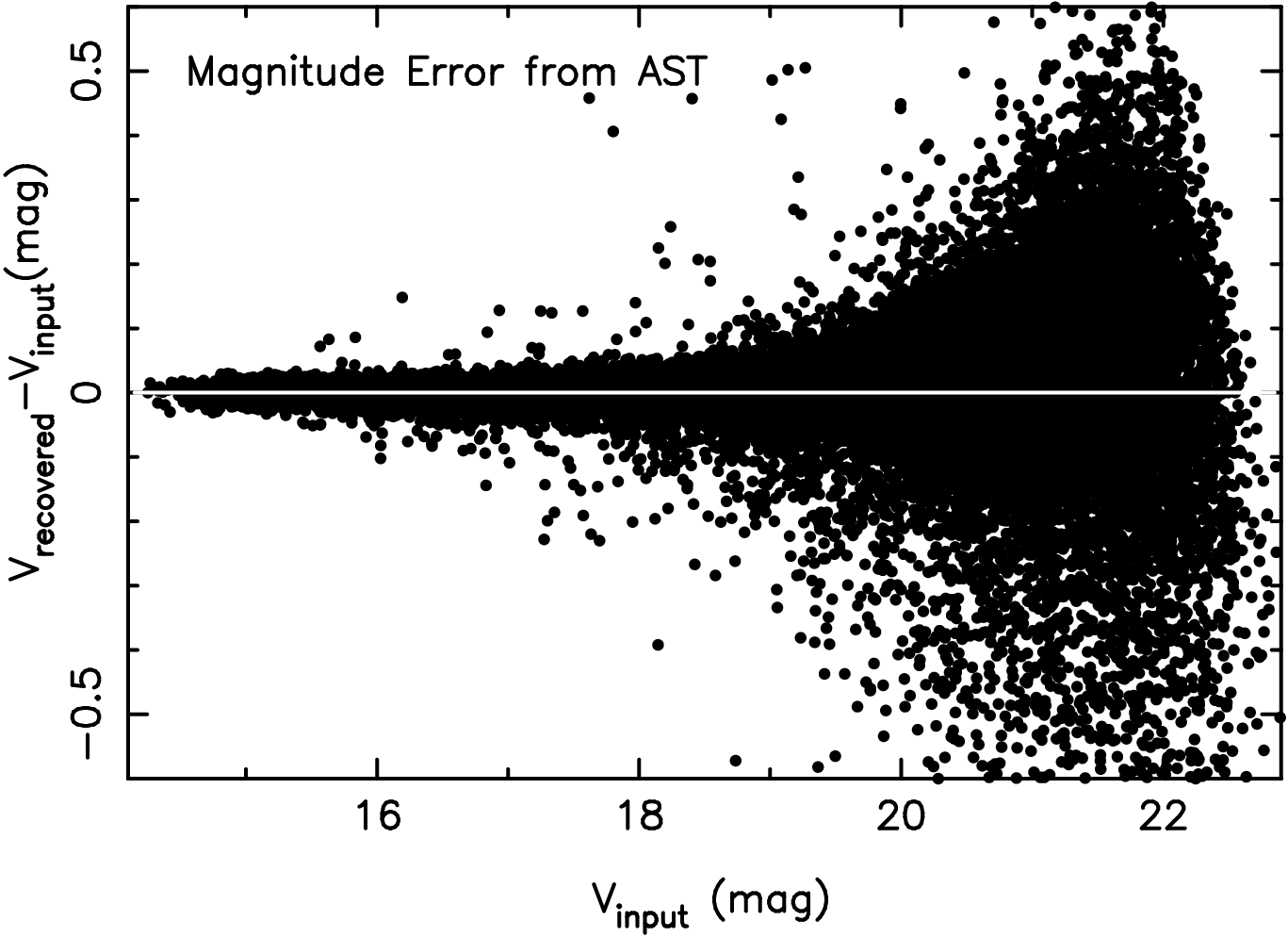}
\caption{Magnitude error as a function of magnitude for cluster NGC1261. Left and right panels are for $B$ and $V$ magnitudes, respectively.}
\end{figure}

\begin{figure} %Fig 7.
\centering
\includegraphics[angle=-90,width=0.45\textwidth]{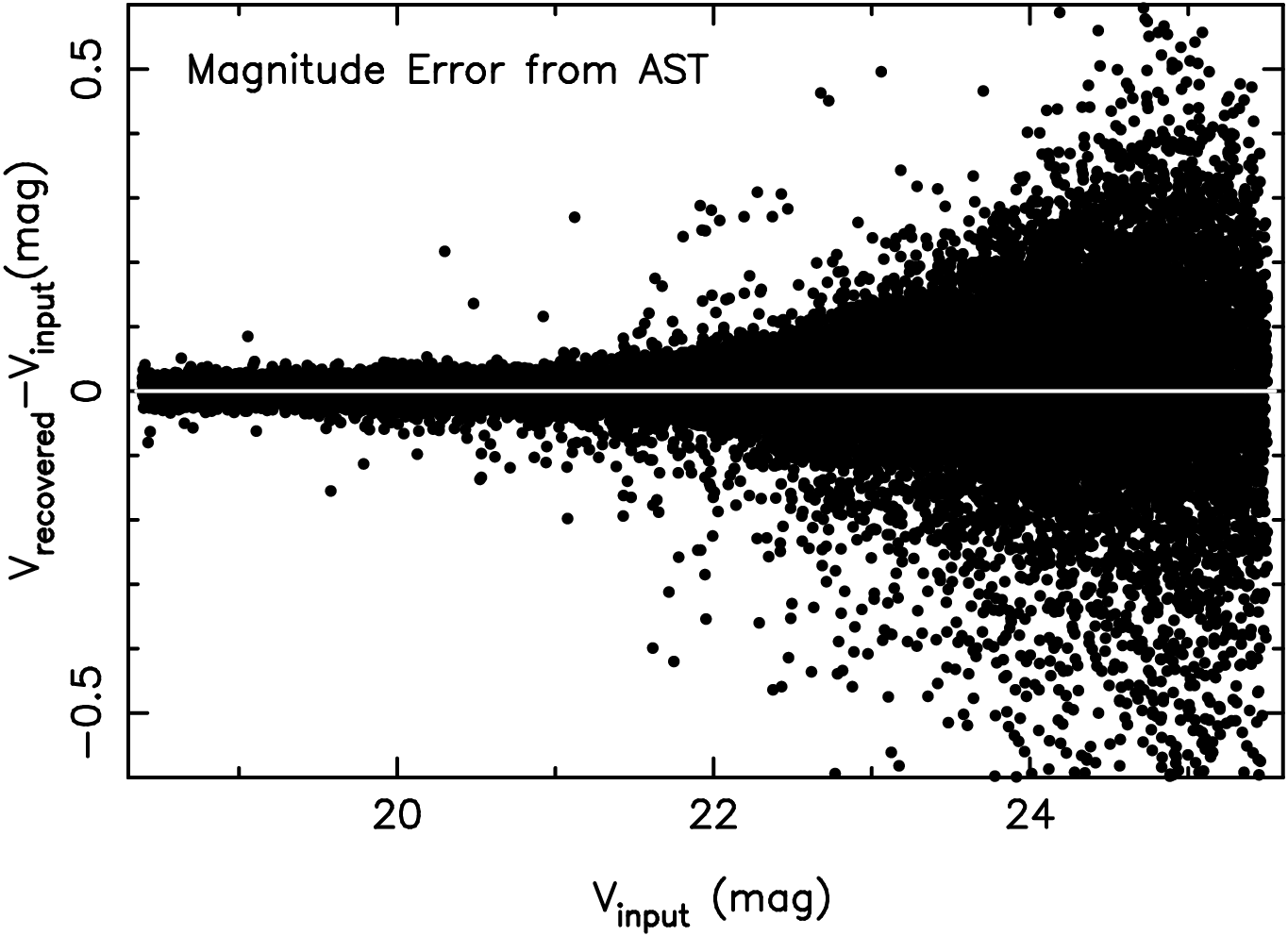}
\includegraphics[angle=-90,width=0.45\textwidth]{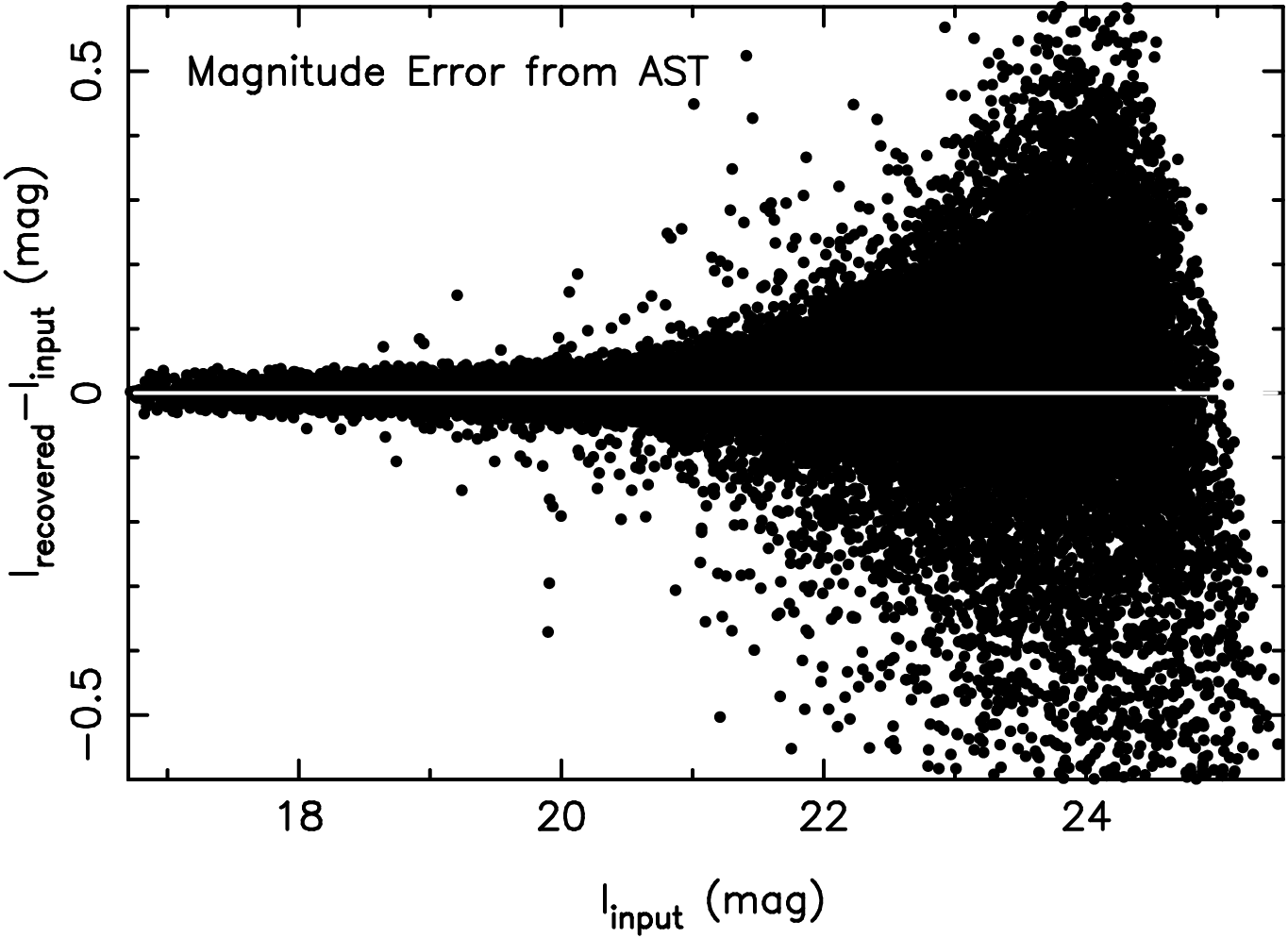}
\caption{Similar to Fig. 1, but for $V$ and $I$ magnitudes of cluster NGC2257.}
\end{figure}

\begin{figure} %Fig 8.
\centering
\includegraphics[angle=-90,width=0.45\textwidth]{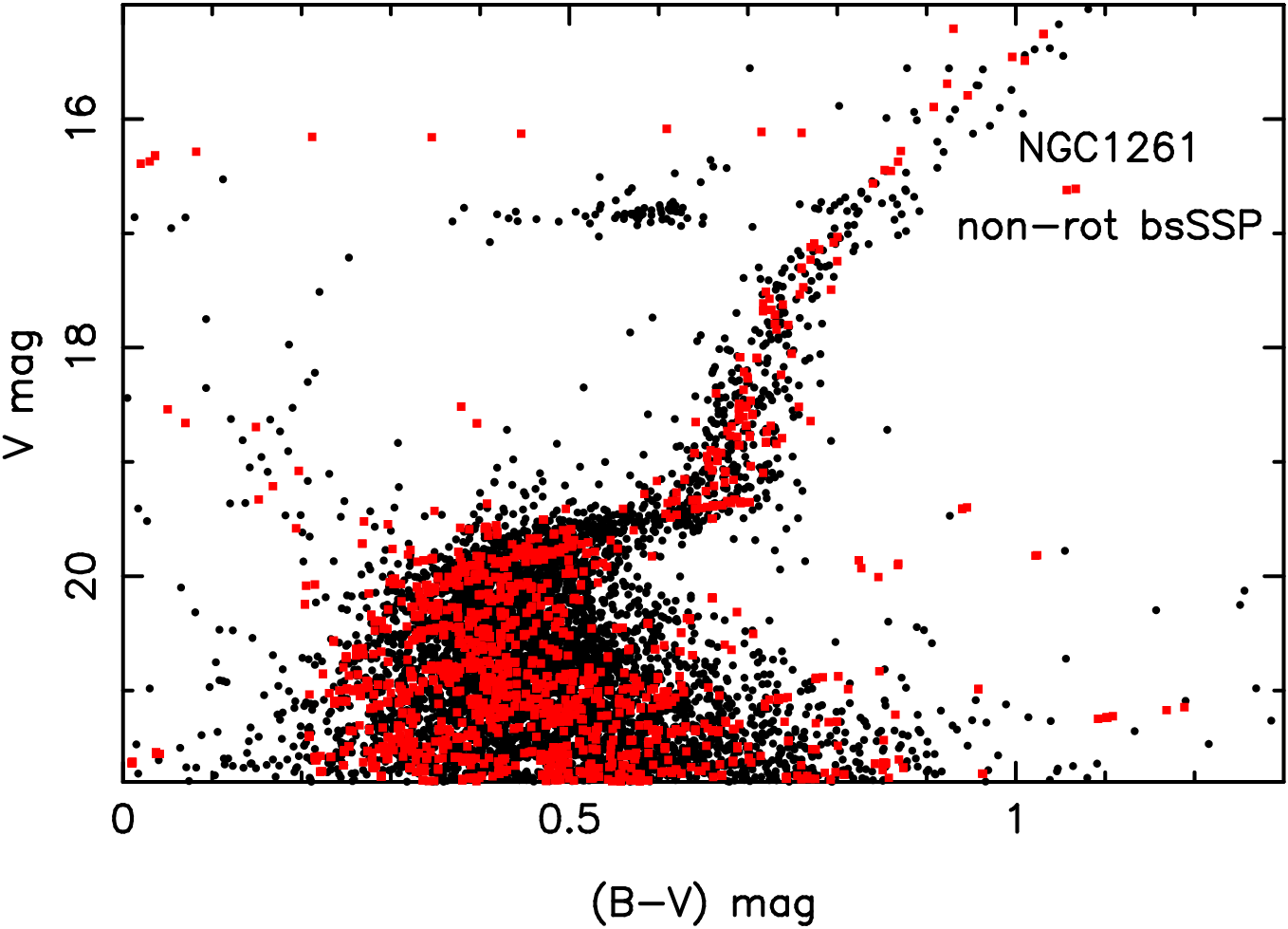}
\includegraphics[angle=-90,width=0.45\textwidth]{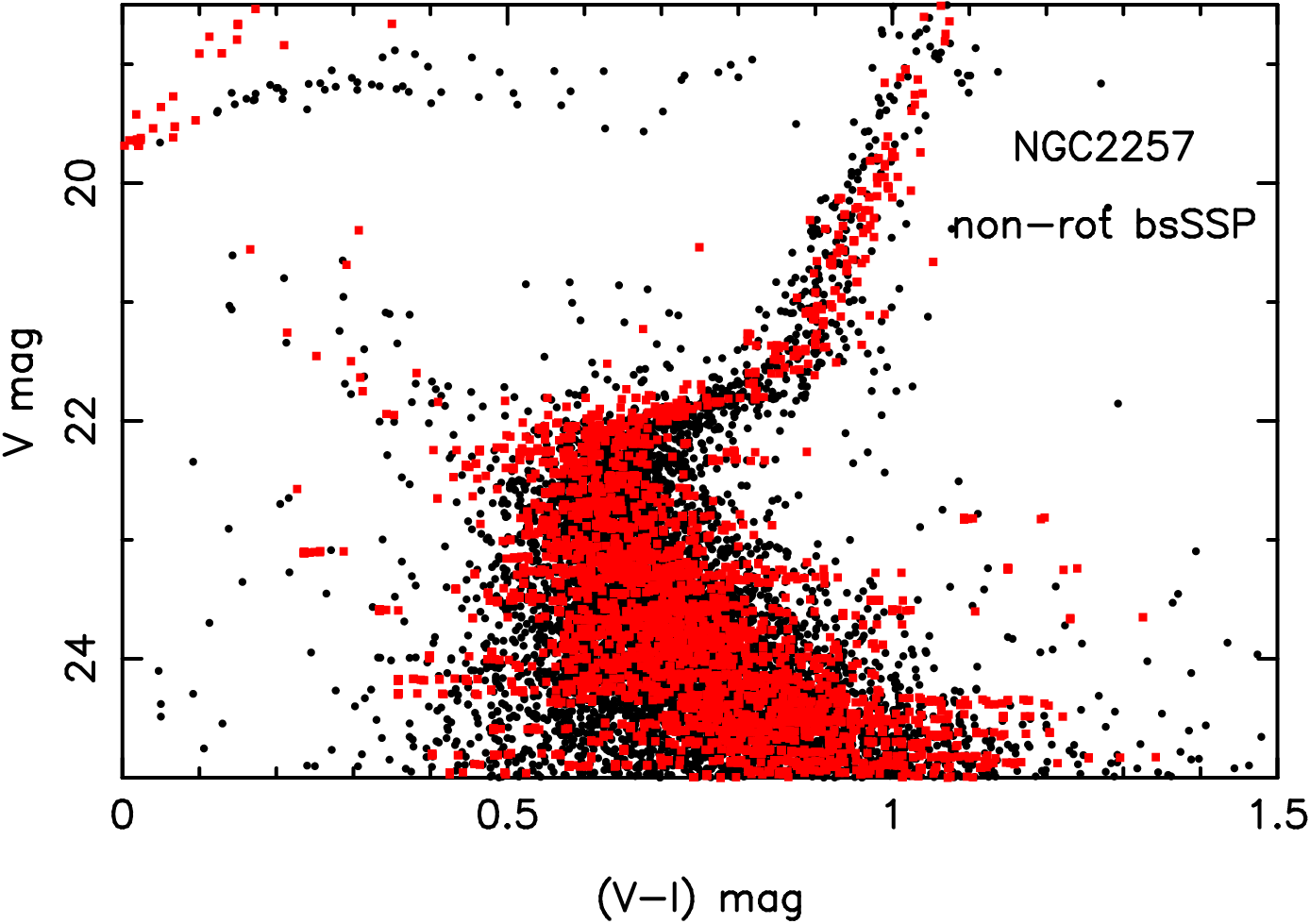}
\caption{Comparison of observed and best-fit CMDs for star clusters NGC1261 and NGC2257.
Black points and red squares are for observed and best-fit CMDs, respectively.
The best-fit parameters of NGC1261 are $Z$ = 0.0003, $m-M$=16.30, $E(B-V)$ = 0.00, age =  13.5\,Gyr, and $f_{\rm b}$ = 0.6.
Those for NGC2257 are $Z$ = 0.0003, $m-M$=18.45, $E(V-I)$ = 0.06, age =  13.6\,Gyr, and $f_{\rm b}$ = 0.8.}
\end{figure}

\section{Results from different models}
In this section we use eight kinds of stellar population models to study the observed CMD of
NGC1651 and compare the results.
These models are different from each other by binary fraction, fraction of rotational stars, star formation types (single or multiple bursts), and star formation histories.
The eight types of models and their abbreviations are listed in Table 1. They are used to derive the best-fit model based on $WAD$ fitting.
Note that the observed CMD possibly contains a few field stars, e.g., those below the red giant branch, but they did not affect the best-fit
result as they enlarge the $WAD$ values of all models simultaneously.

\begin{table}
 \caption{Model numbers (No.) and abbreviations of eight types of stellar population models.
 Fractions of binaries and rotational stars can be set to any value.}
 \label{symbols}
 \begin{tabular}{lll}
  \hline
  No. &Abbreviation   &Model\\
 \hline
1 &non-rot ssSSP &non-rotational single star simple stellar population\\
2 &non-rot bsSSP &non-rotational binary star simple stellar population\\
3 &non-rot ssCSP &non-rotational single star composite stellar population\\
4 &non-rot bsCSP &non-rotational binary star composite stellar population\\
5 &std-rot ssSSP &rotational single star simple stellar population\\
6 &std-rot bsSSP &rotational binary star simple stellar population\\
7 &std-rot ssCSP &rotational single star composite stellar population\\
8 &std-rot bsCSP &rotational binary star composite stellar population\\
  \hline
 \end{tabular}
 \end{table}

In CMD fitting, the magnitude uncertainties of a simulated star
are taken from those of AST stars with similar (difference less than 0.05\,mag) magnitudes.
Distance modulus, color excess, binary fraction, rotational star
fraction, and star formation mode of a cluster are determined
simultaneously via the $WAD$ fitting method, because it was shown to be reliable.
A new CMD fitting code, i.e., ``Binary Star to Fit for CMD" (\emph{BS2fit for CMD}),
which was developed by Dr. Zhongmu Li, is used for the study.
The code is a part of research project \emph{Binary Star to Fit (BS2fit)},
and it will be available to the public in the future.
We use five star formation modes to describe the SFH of clusters in a simple way.
The five modes are uniform, linearly increasing, linearly decreasing, Gaussian,
and decreasing-increasing (``V'' mode) formation modes.
In each mode, the fraction of stars of each single burst is given as a function of stellar age.

The ranges and steps of parameters for the final run of CMD fitting of NGC1651 are shown in Table 2.
The values are estimated by some runs that took larger ranges and steps for these parameters in advance.

\begin{table}
 \caption{Parameter ranges and steps for last run of CMD fitting of NGC1651.
 $N_{\rm sf}$, $M_{\rm sf}$, $f_{\rm bin}$, and $f_{\rm rot}$ are
 number of star bursts, mode of star formation, binary fraction, and rotational star fraction, respectively.
 $M_{\rm sf}$ from 1 to 5 means uniform, linearly increasing, linearly decreasing, Gaussian and decreasing-increasing modes, respectively.}
 \label{symbols}
 \begin{tabular}{llll}
  \hline
  parameter   &range                &step     &unit\\
 \hline
 $Z$          & 0.008               &         &   \\
 $(m-M)_0$      &19.10--19.20       &0.01     &mag\\
 $E(V-I)$     & 0.00-- 0.25         &0.01     &mag\\
 $N_{\rm sf}$ &    1-- 5            &1        &   \\
 $M_{\rm sf}$ &    1-- 5            &1        &   \\

 Age          & 1.0 -- 2.5          &0.1      &Gyr\\
 $f_{\rm bin}$& 0.0 -- 0.8          &0.1      &   \\
 $f_{\rm rot}$& 0.0 --100.0         &0.1      &   \\
  \hline
 \end{tabular}
 \end{table}

 Table 3 shows the best-fit results from various models. As we see, when a relationship of $A_{\rm v} = 2.253*E(V-I)$ is adopted, the best-fit true distance modulus of cluster NGC1651 is possibly between 18.68 and 18.70,
 which is consistent with previous results (18.5 -- 18.7, see
 \citealt{Mould1986,Sarajedini2002}). The SSP-fit age (1.5 -- 1.6\,Gyr) of this cluster is
 different from \cite{Sarajedini2002} ($\sim$ 1.8\,Gyr), but is
 similar to that of \cite{Mould1997} and \cite{Mould1986} ($\sim$ 1.6\,Gyr).
 The total color excess $E(V-I)$ is about 0.19 \,mag.
 In addition, this cluster possibly contain about
 50 -- 60 per cent binaries. Note that
 `binaries' here mean the ones with orbital period less than 100\,yr today.
 Because interacting binaries are only a part of such binaries,
 binary fraction ($f_{\rm bin}$) shown by this work is different from some other works (e.g., \citealt{davis2008}),
 in which binary fraction usually means fraction of interacting binaries.

Panel $(a)$ of Fig. 9 shows the best-fit CMD from the traditional kind of model, i.e., non-rot ssSSP. Complementary information is also present in Fig. 10.
As we see, there are obvious differences between the best-fit (red) and observed (black) CMDs.
The theoretical model does not reproduce extended MS, turn-off, and red clump well.
These three parts are labelled as parts 1, 2 and 3, respectively, to help readability of the figures. Note, RC of a non-rot ssSSP is like a simple curve, as shown by red points.

Then Panel $(b)$ shows the result from another kind of simple stellar population model,
in which the effects of binaries have been taken into account.
We find that although broad MS and blue stragglers
are reproduced, there is significant difference between the observed
and best-fit CMDs, in particular for the turn-off (``1'') and red clump (``3'')parts.
This indicates that binaries are not the main cause of eMSTO (see also the review of
\citealt{Girardi13}) and eRC, although this contribute a few (about 6) stars to eMSTO.
It suggests that the stellar population of NGC1651 is not an SSP without rotational stars.
Our result is therefore different from that of \cite{Yang11}, which takes binaries as a main reason for eMSTO because an overlarge star number (50\,000 binaries) and a fixed binary fraction was used. In their work, the number of binaries are considerably overestimated.

After that panel $(c)$ shows the result from non-rot ssCSP fits. In the best-fit model, single stars formed by a few bursts, in fact all stars, have the same metallicity.
It is shown that the best-fit model reproduces eMSTO and eRC, but it does not reproduce well the broad MS (``2'') and blue stragglers (upper left).
This implies that multiple star bursts may be the reason for eMSTO and eRC.

Next, panel $(d)$ shows the result of a similar model, but taking binaries into account.
We are shown that the best-fit population (non-rot bsCSP) fits the observed CMD much better than previously tested models.
The main observational features including eMSTO, broad MS, eRC, and blue stragglers are reproduced here. The $WAD$ of this model is much less than that of non-rot ssSSP, non-rot bsSSP, and non-rot ssCSP models (Table 3). Thus, among all non-rotational models, a bsCSP (parameters can be found in Table 3) is the best explanation for the observed CMD of NGC1651.

If we use another kind of simple stellar population model, which takes the effects of stellar rotation into account,
some new results are shown. Panels $(e)$ to $(g)$ give the comparison of best-fit and observed CMDs.

Panel $(e)$ shows the result from std-rot ssSSP fits.
In the best-fit model, all stars
are found to be rotators. Such models
can reproduce the eMSTO part (``1''), but cannot reproduce broad MS (``2''), eRC, and blue stragglers, as we can see.
According to this figure, there is an obvious difference between fitted and observed CMDs. The stellar population of NGC1651 is not likely to be a std-rot ssSSP.

When we use rotational binary population models (std-rot bsSSP) for CMD fitting, the result seems obviously better. Although the best-fit population is a simple one that formed its stars via a star burst, all observed features are reproduced by it, and its $WAD$ is small, at only 0.00227. We see that stellar rotation can lead to eMSTO and eRC, because it changes the luminosity, surface temperature, and evolutionary speed of stars. One can read panel $(f)$ for a detailed comparison of best-fit and observed CMDs. Comparing the results of non-rot bsCSP and std-rot bsSSP fits (panels $d$ and $f$) shows that both simple and composite stellar population models can explain the CMD of NGC1651. Therefore, it is not clear whether the eMSTO and eRC result from an age spread (i.e., multiple bursts), as the combination of stellar rotation and binarity can also lead to eMSTO and eRC.

Because it is possible that a cluster contains a fraction of rotational stars with different ages, we also test such models. Panel $(g)$ compares the best-fit std-rot ssCSP model and the observation.
We find that broad MS (``2'') and blue stragglers are not well-fitted, while we find that std-rot bsCSP models fit best to the observed CMD (panel $h$) among eight kinds of models. This suggests that the stellar population of NGC1651 may contain 50 per cent binaries and 70 per cent rotators. The ages of stars can vary from 1.4 to 1.8\,Gyr.

In order to make the comparison between fitted and observed CMDs clearer,
Fig. 11 shows the difference between observed CMD and best-fit models (those shown in Fig. 9).
By comparing two panels at the same line (e.g., panels $a$ and $b$), we can conclude from the figure that stellar population models with binaries fit to the observed CMD better than those without binaries.
Meanwhile the comparisons of panels $b$ and $f$, and panels $d$ and $h$ indicate that models with rotational stars reproduce the observed CMD features better than those without rotators. This suggests that both binaries and rotational stars possibly contribute to the special CMD of NGC1651.
Because the fit of eMSTO part from the std-rot bsSSP (panel $f$) is better than from non-rot bsCSP (panel $d$), while it is not as good as that from std-rot bsCSP (panel $h$),
the eMSTO of NGC1651 could be caused by not only age spread, but also stellar rotation and binarity. It is also possible that the observed CMD result from all the three reasons. There is a degeneracy between their effects on the CMD.
In addition, we find from panels $(f)$ and $(h)$ of Fig. 10 that there are only a few grids with differences larger than 0.4 $f0_{\rm diff}$.
This implies that most parts of observed CMD can be reproduced well by both the std-rot bsSSP and std-rot bsCSP models.

Finally, we check the color and magnitude distributions of observed and fitted CMDs in
Figs. 11 and 12. It can be seen from Fig. 11 that for the part near eMSTO ($V-I$ = 0.4 $\sim$ 0.6\,mag), composite stellar population models
(non-rot bsCSP and std-rot bsCSP) have color distributions closer
to the observed result, compared to simple stellar population models (non-rot bsSSP and std-rot bsSSP). This is in agreement with the result of Fig. 9.
At the same time, the distribution of magnitude has a similar trend for all models, as shown in Fig. 12 ($V$ = 20.5 $\sim$ 22.0\,mag).
We can thereby conclude that composite population models can better reproduce the eMSTO structure.

From the comparison in Figs. 9 to 12, we find
that all models have some difference from the observed CMD of NGC1651,
because all $WAD$ values are greater than zero and there is
obvious difference in the color and magnitude distributions between the observed and fitted CMDs.
A main part of this difference may result from the uncertainties in IMF
and distribution of initial binary separation of stellar populations.
Meanwhile, the $WAD$ of std-rot bsSSP is close to that of
non-rot bsCSP. This implies that the stellar
population of NGC1651 is equally possible to be a simple stellar
population of rotating stars and a composite population
without rotators. When comparing the $WAD$ values of best-fit std-rot bsSSP
and std-rot bsCSP models, we find a small difference of
0.0002188. This is similar to the uncertainty in CMD synthesis,
which can be caused by the uncertainties in rotation rate
distribution, and in the treatment of the effect of stellar rotation on
magnitudes of stars. If we change the distribution of stellar
rotation rate, the $WAD$ of std-rot bsSSP can possibly be smaller.
Therefore, it is difficult to reject the model of
bsSSP with rotating stars from our results. In this case, to study
whether clusters like NGC1651 contain a large number of rotational
stars is crucial for understanding the stellar populations of such
clusters.

\begin{table}
 \caption{Best-fit parameters of NGC1651 from various models.
 The meanings of models corresponds to Table 1 via model numbers, and the units
 of parameters are the same as in Table 2.
 The value of $(m-M)_0$ has corrected for the galactic extinction.
 $WAD$ is the difference of star fraction in an effective grid of 0.047\,mag for color and 0.020\,mag for magnitude (total effective grid number is about 2230).
 The less the $WAD$, the better the CMD fit.}
 \label{symbols}
 \begin{tabular}{llllllllll}
  \hline
  Parameter   &Model 1         &Model 2         &Model 3       &Model 4        &Model 5         &Model 6         &Model 7       &Model 8      \\
 \hline
 $Z$          &0.008           &0.008           &0.008         &0.008          &0.008           &0.008           &0.008         &0.008        \\
 $(m-M)_0$    &18.72           &18.70           &18.70         &18.68          &18.71           &18.70           &18.69         &18.70        \\
 $E(V-I)$     & 0.19           &0.19            &0.19          &0.18           &0.19            &0.18            &0.18          &0.19         \\
 $N_{\rm sf}$ & 1              &1               &6             &6              &1               &1               &6             &5            \\
 $M_{\rm sf}$ &--              &--              &5             &4              &--              &--              &5             &3            \\
 Age (Gyr)    &1.5             &1.6             &1.4--1.9      &1.4--1.8       &1.5             &1.5             &1.4--1.9      &1.4--1.8     \\
 $f_{\rm bin}$&0               &0.5             &0             &0.5            &0               &0.6             &0             &0.5          \\
 $f_{\rm rot}$&0               &0               &0             &0              &1.0             &1.0             &0.3           &0.7          \\
 $WAD$x1000   &3.7381          &2.4689          &3.0112        &\textbf{2.1692}&3.3273          &\textbf{2.2674} &3.2933        &\textbf{2.0486}\\
  \hline
 \end{tabular}
 \end{table}

 \begin{figure} %Fig 9.
\centering
\includegraphics[angle=-90,width=0.4\textwidth]{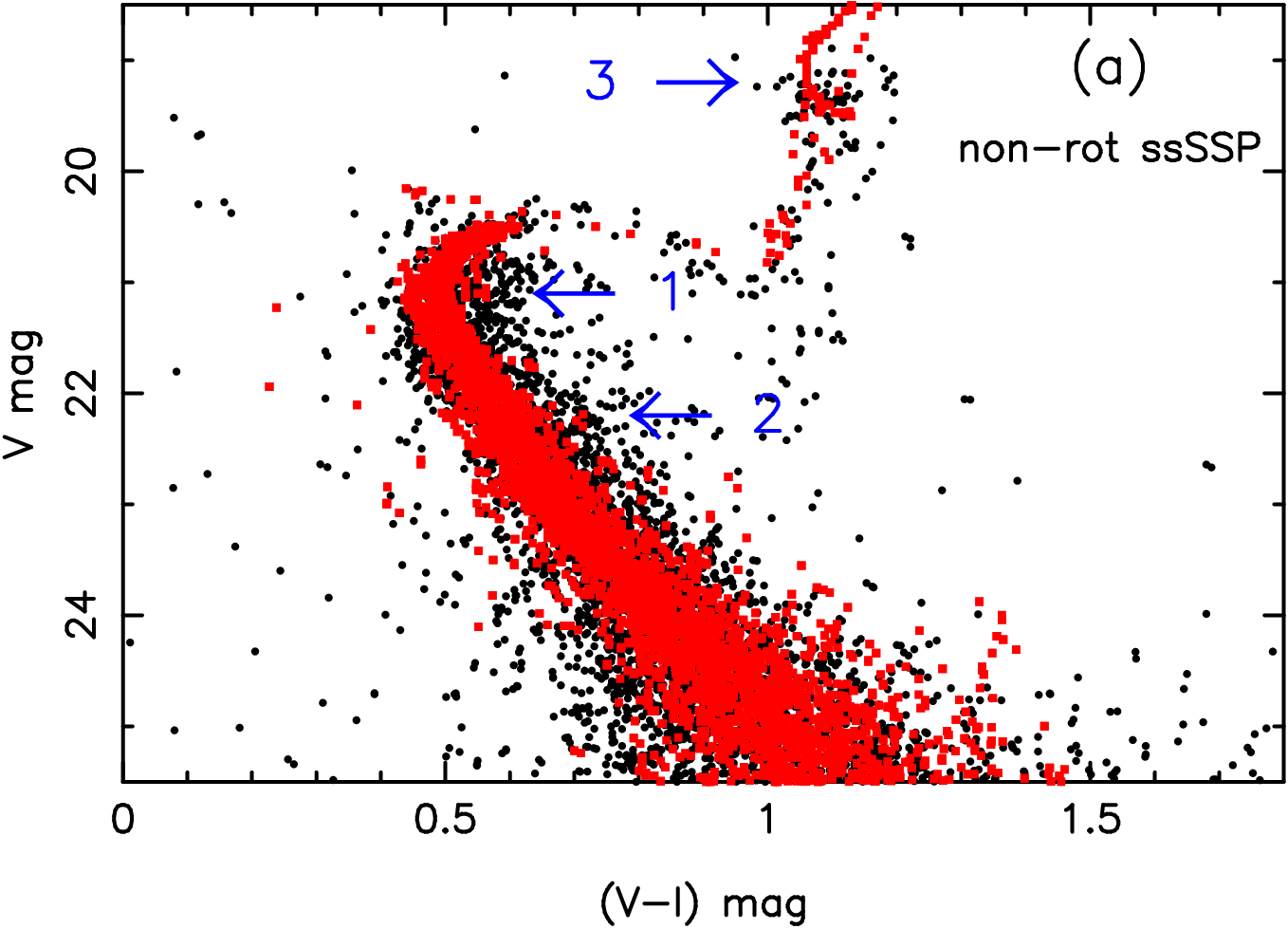}
\includegraphics[angle=-90,width=0.4\textwidth]{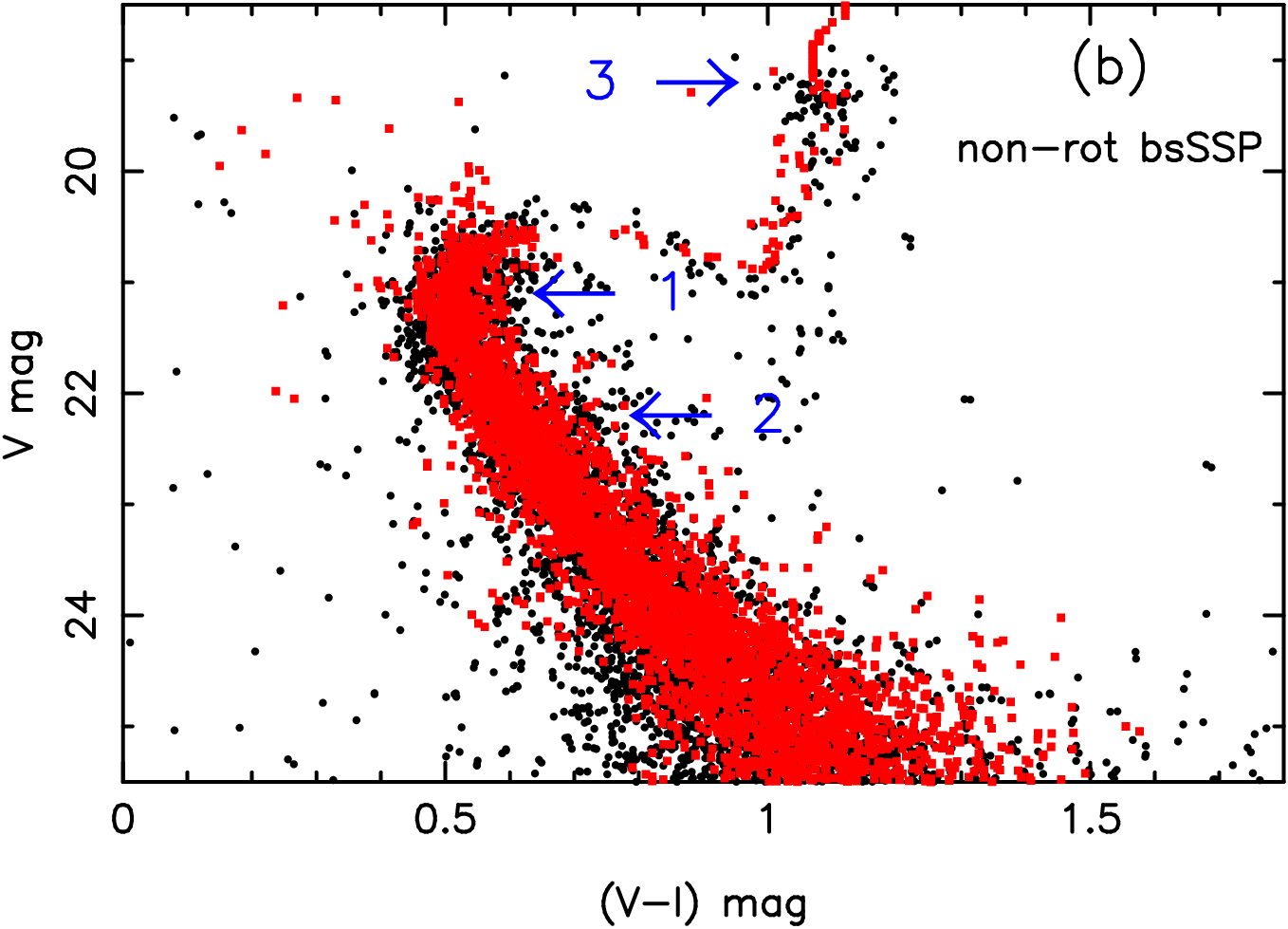}
\includegraphics[angle=-90,width=0.4\textwidth]{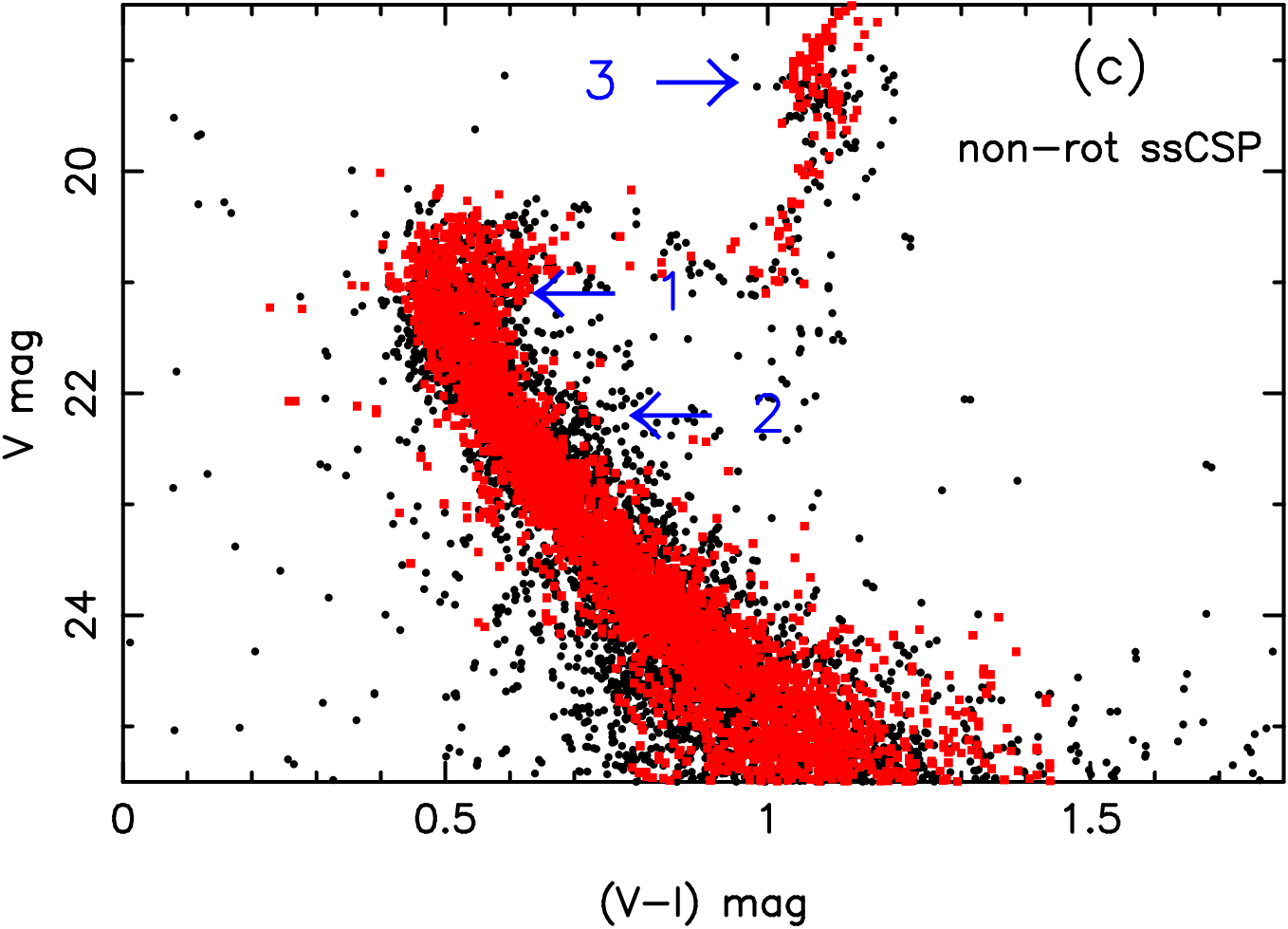}
\includegraphics[angle=-90,width=0.4\textwidth]{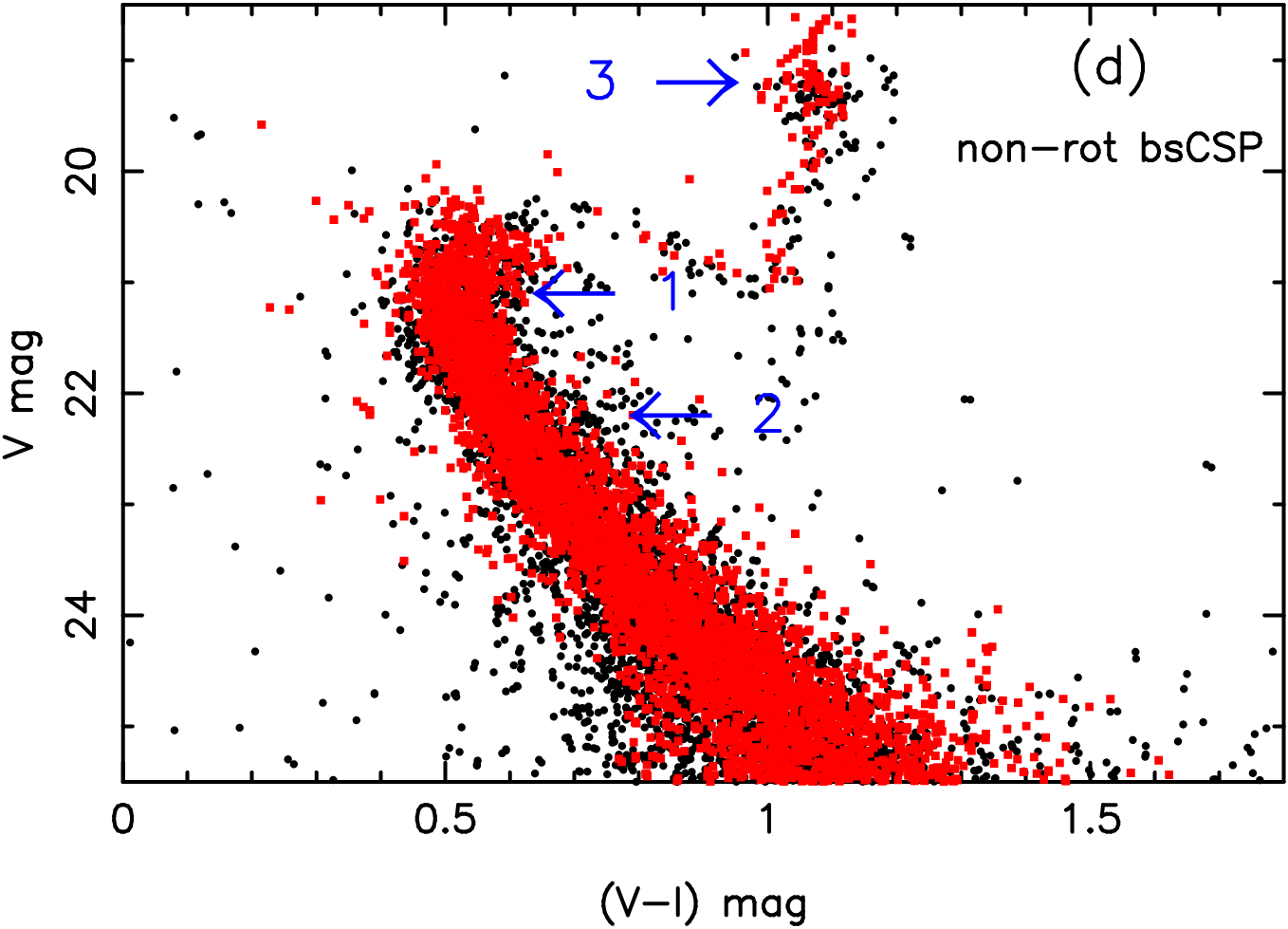}
\includegraphics[angle=-90,width=0.4\textwidth]{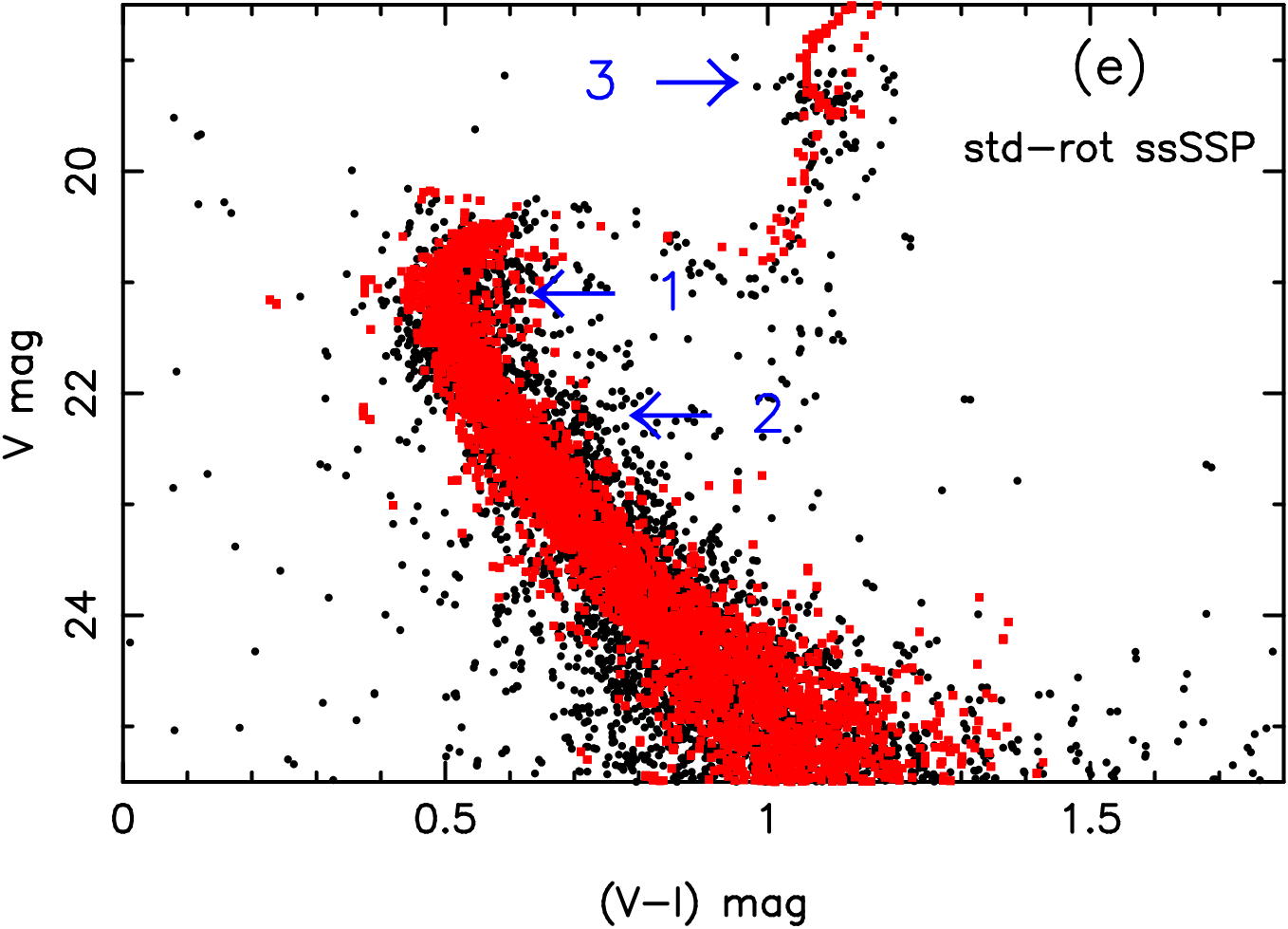}
\includegraphics[angle=-90,width=0.4\textwidth]{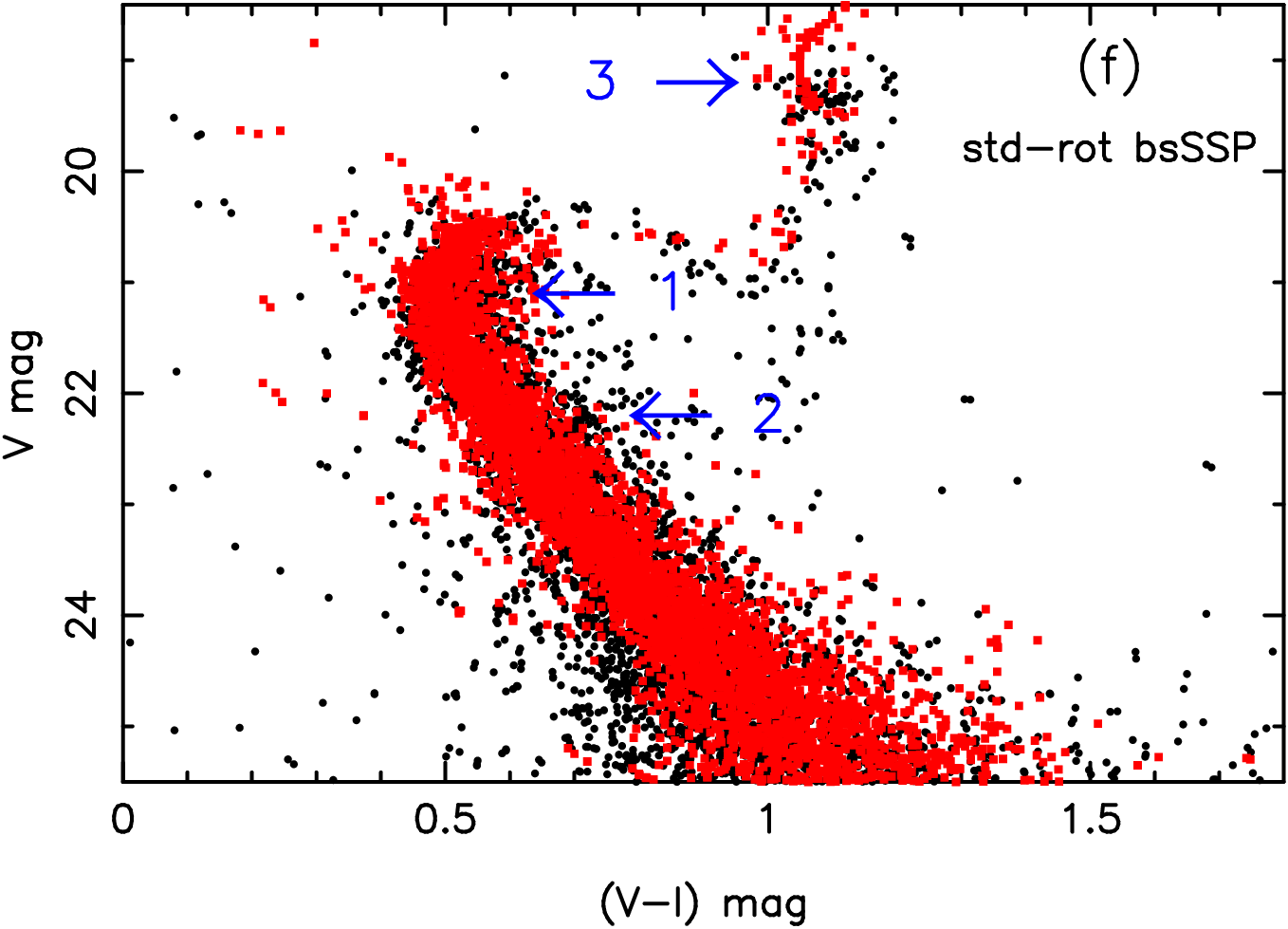}
\includegraphics[angle=-90,width=0.4\textwidth]{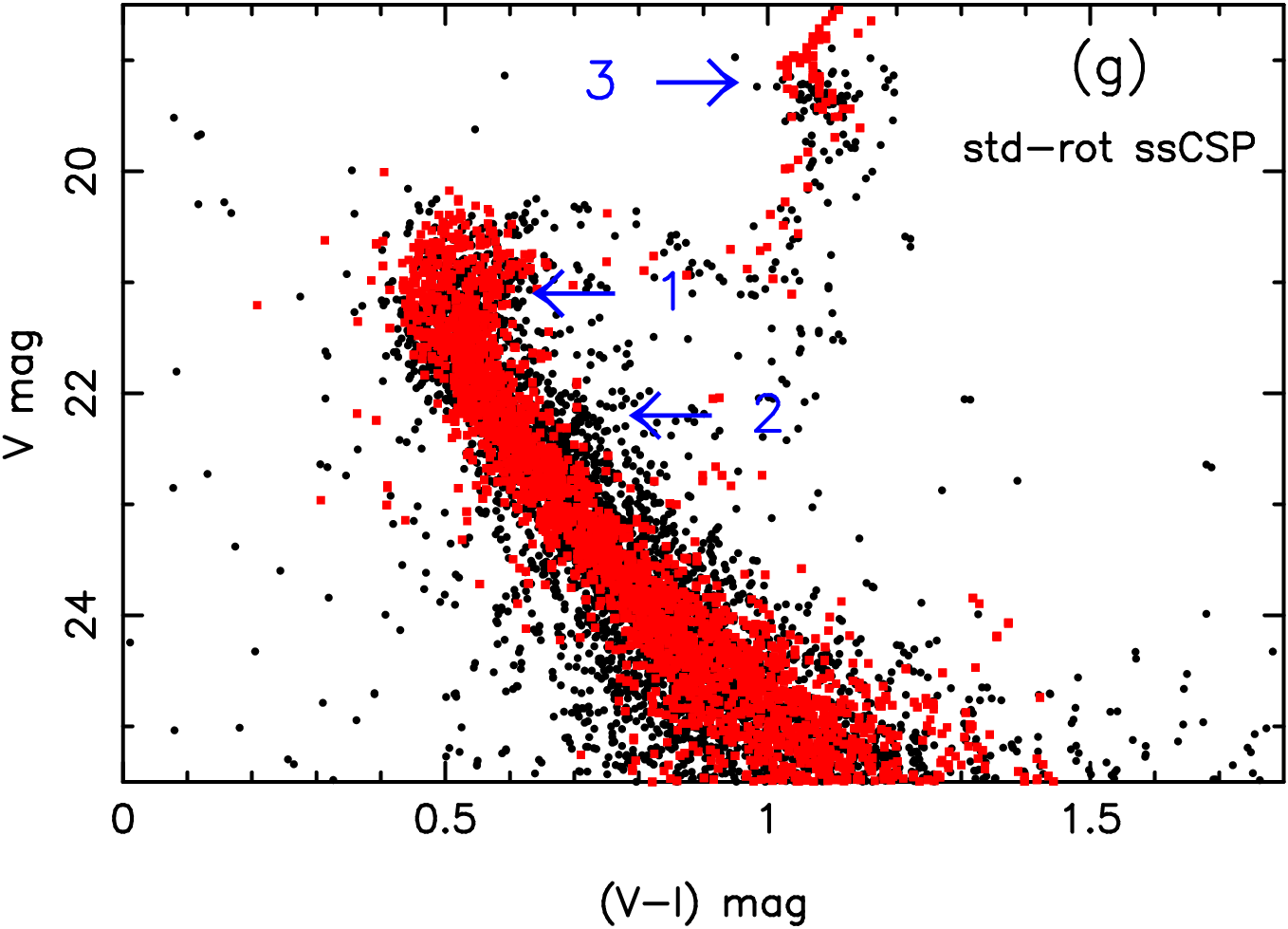}
\includegraphics[angle=-90,width=0.4\textwidth]{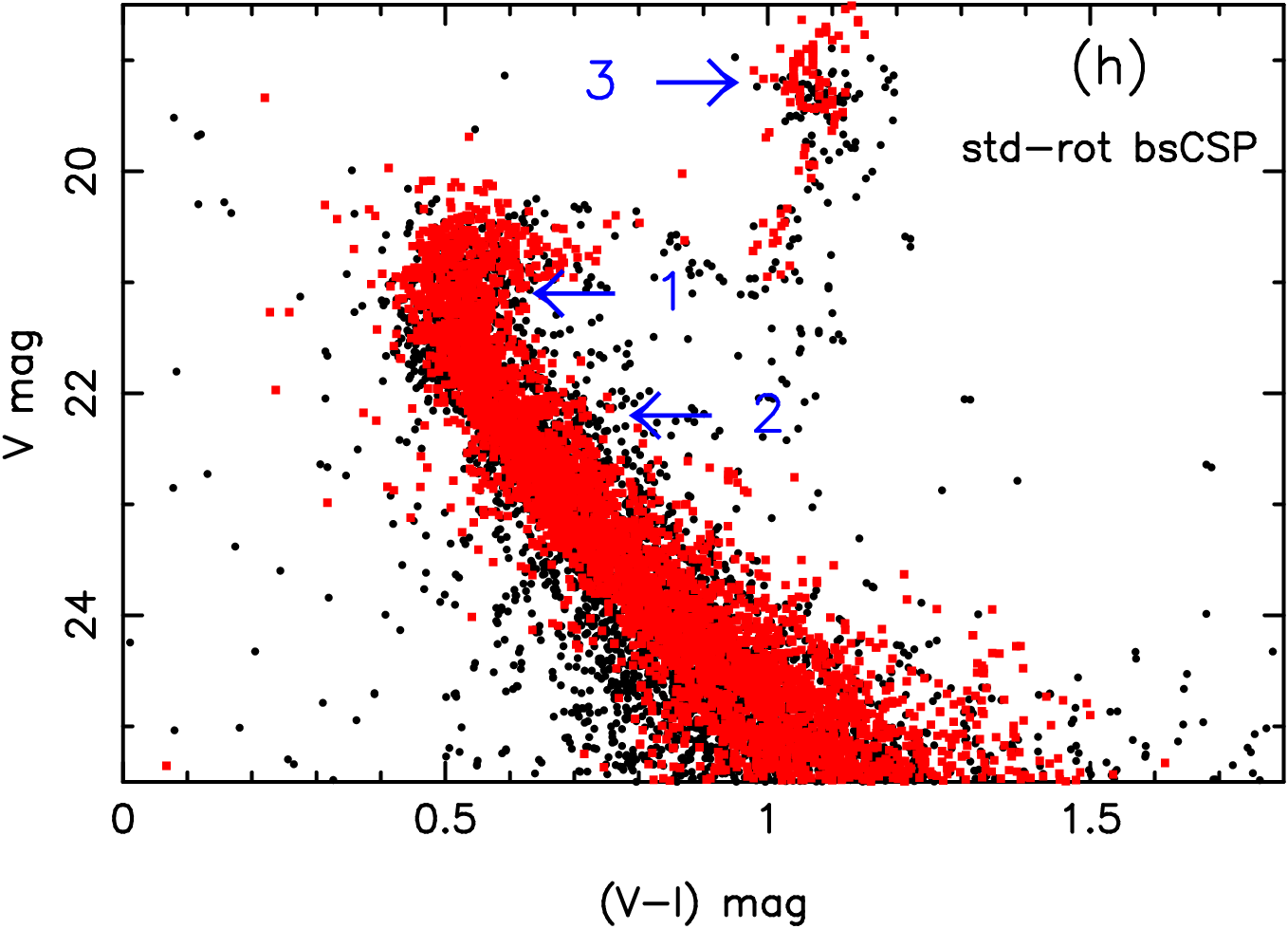}
\caption{Comparison of best-fit CMDs to observed CMD of NGC1651.
Black points and red squares denote observed and synthetic CMDs, respectively.
``std-rot'' and ``non-rot'' indicate populations with and without rotational stars.
``ss'' and ``bs'' respectively mean single and binary star population. ``SSP'' and ``CSP'' denote simple and composite stellar
population. Detailed fitting results can be found in Table 3.
Magnitude errors are randomly generated on the basis of correlations between magnitude and its error (see Fig. 1).}
\end{figure}

 \begin{figure} %Fig 10.
\centering
\includegraphics[angle=-90,width=0.45\textwidth]{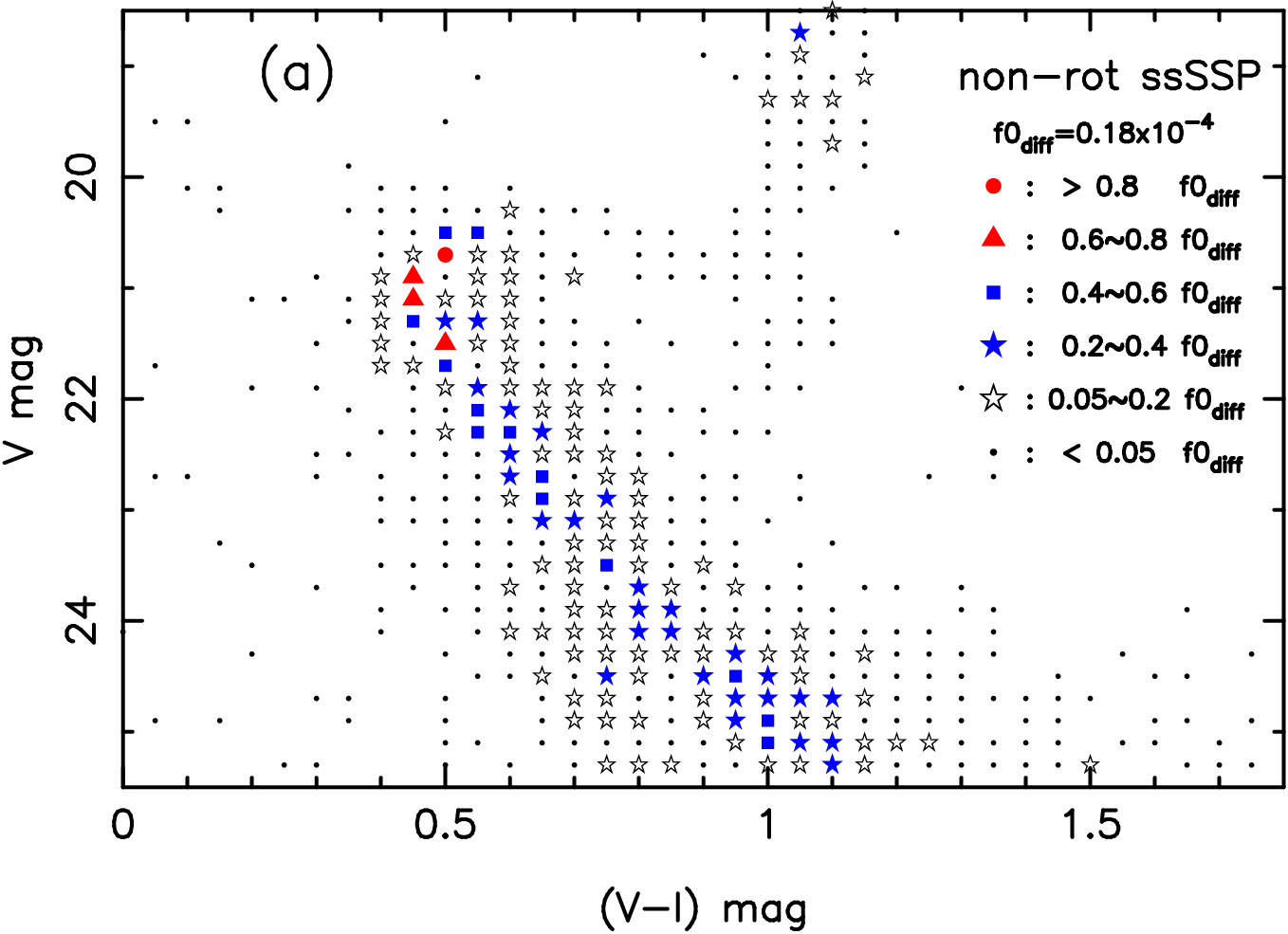}
\includegraphics[angle=-90,width=0.45\textwidth]{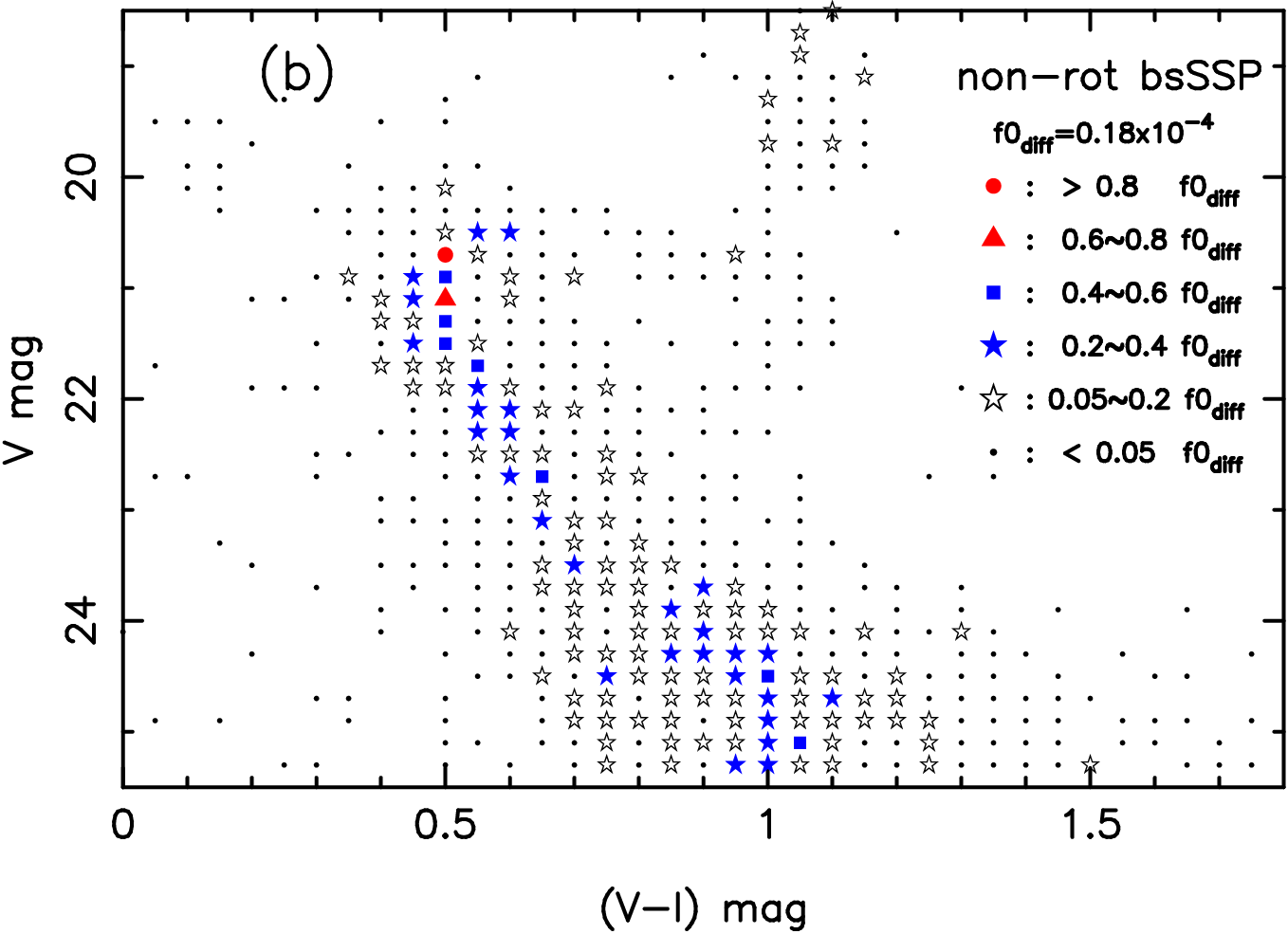}
\includegraphics[angle=-90,width=0.45\textwidth]{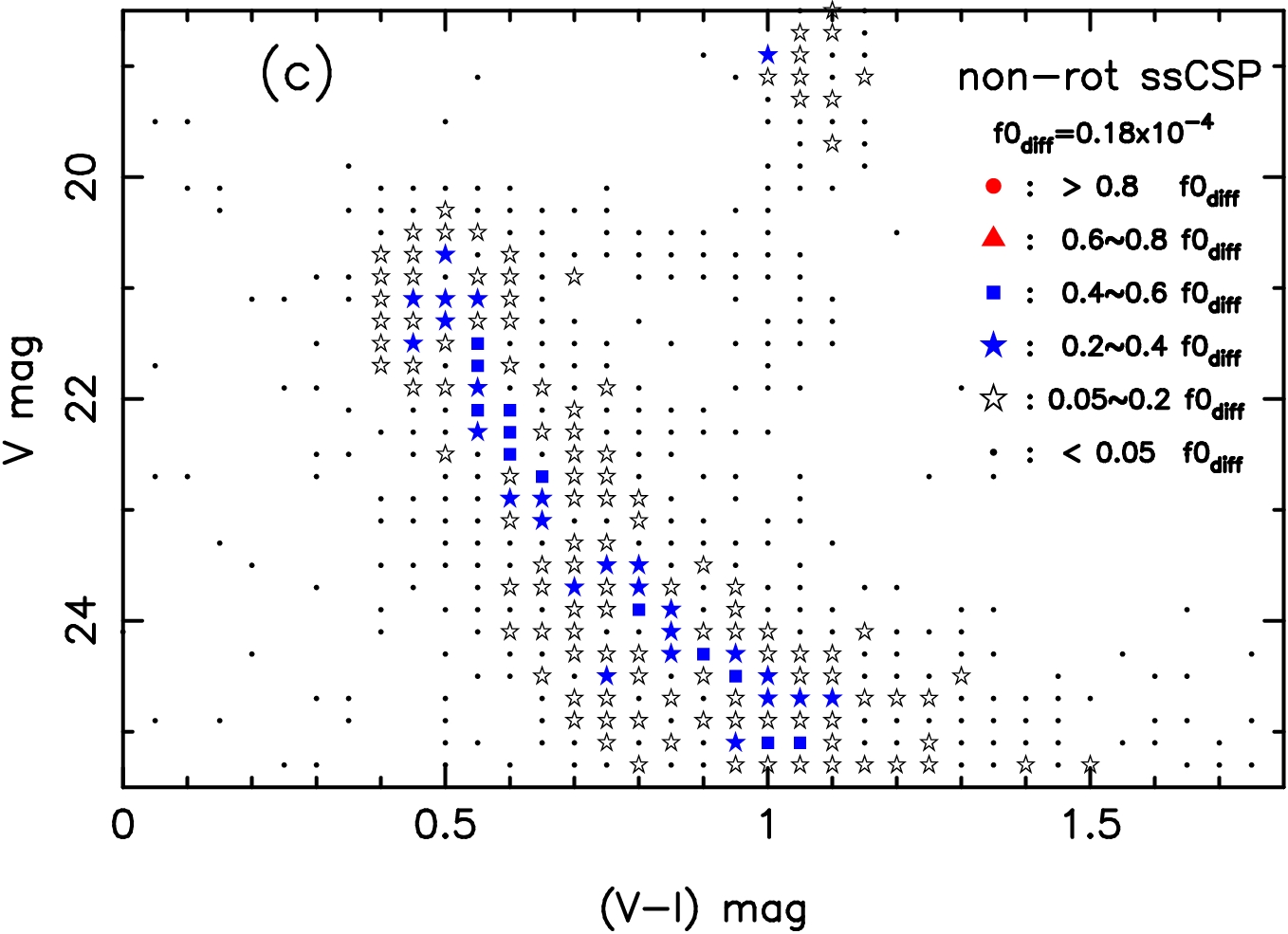}
\includegraphics[angle=-90,width=0.45\textwidth]{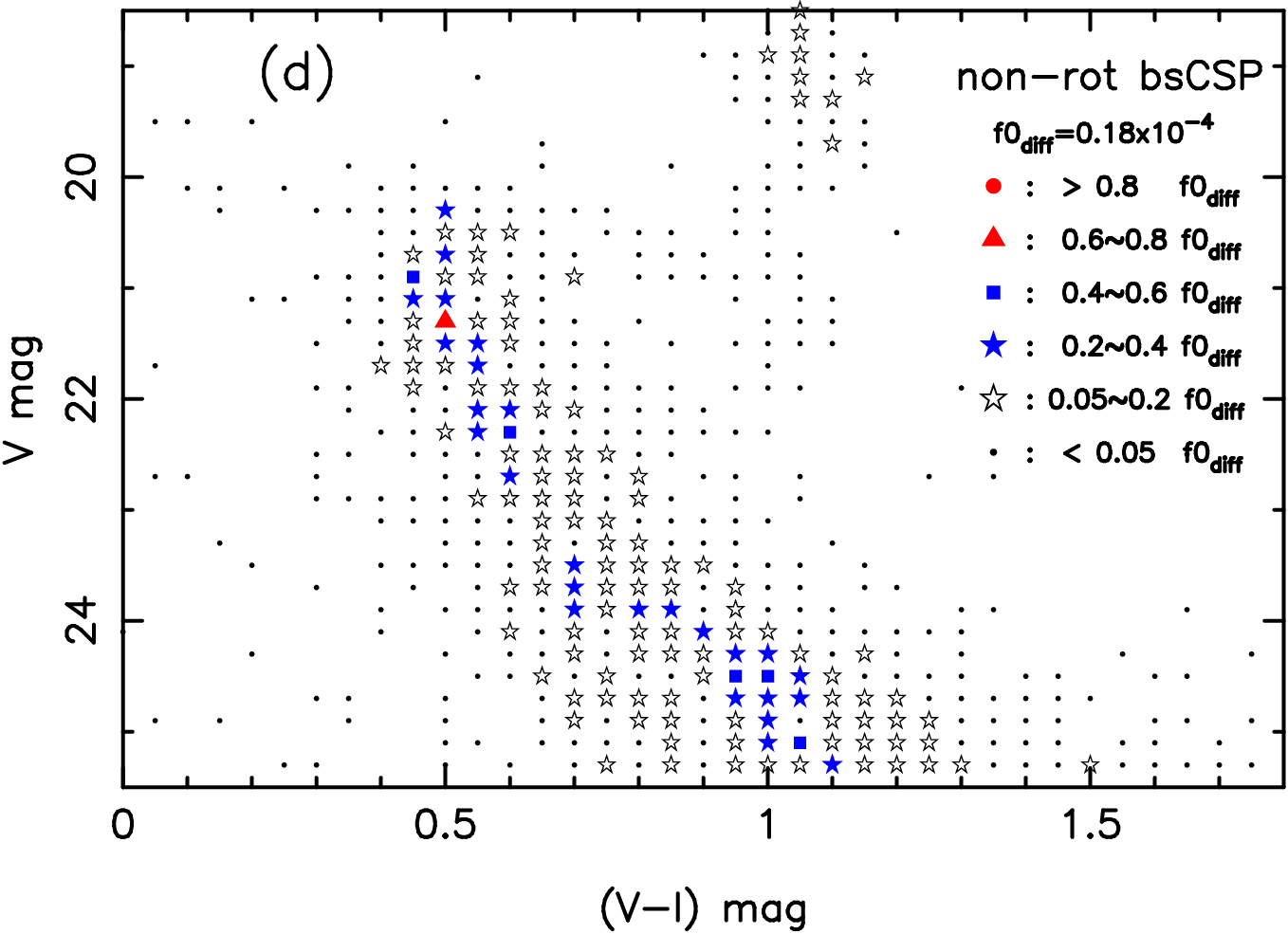}
\includegraphics[angle=-90,width=0.45\textwidth]{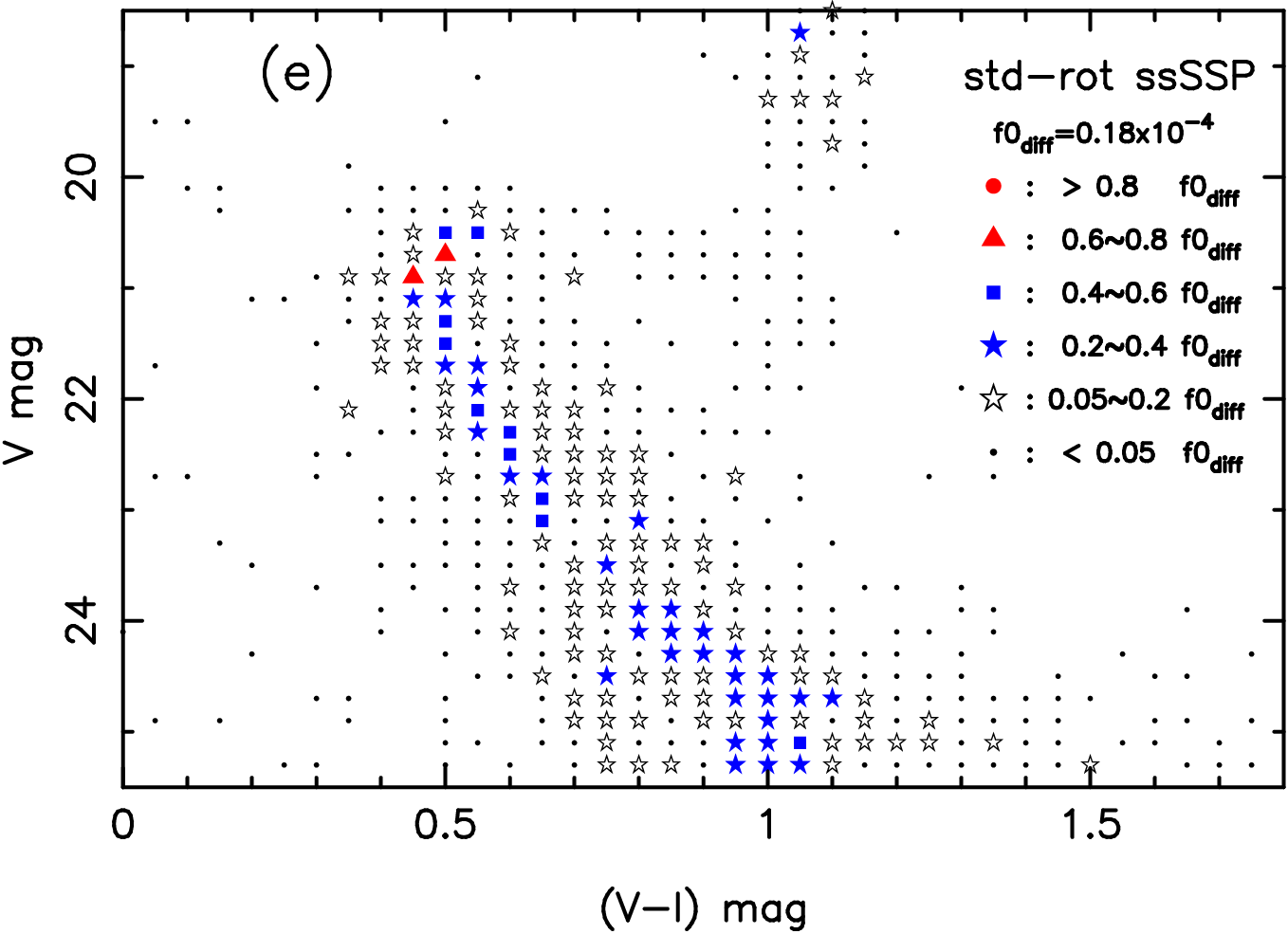}
\includegraphics[angle=-90,width=0.45\textwidth]{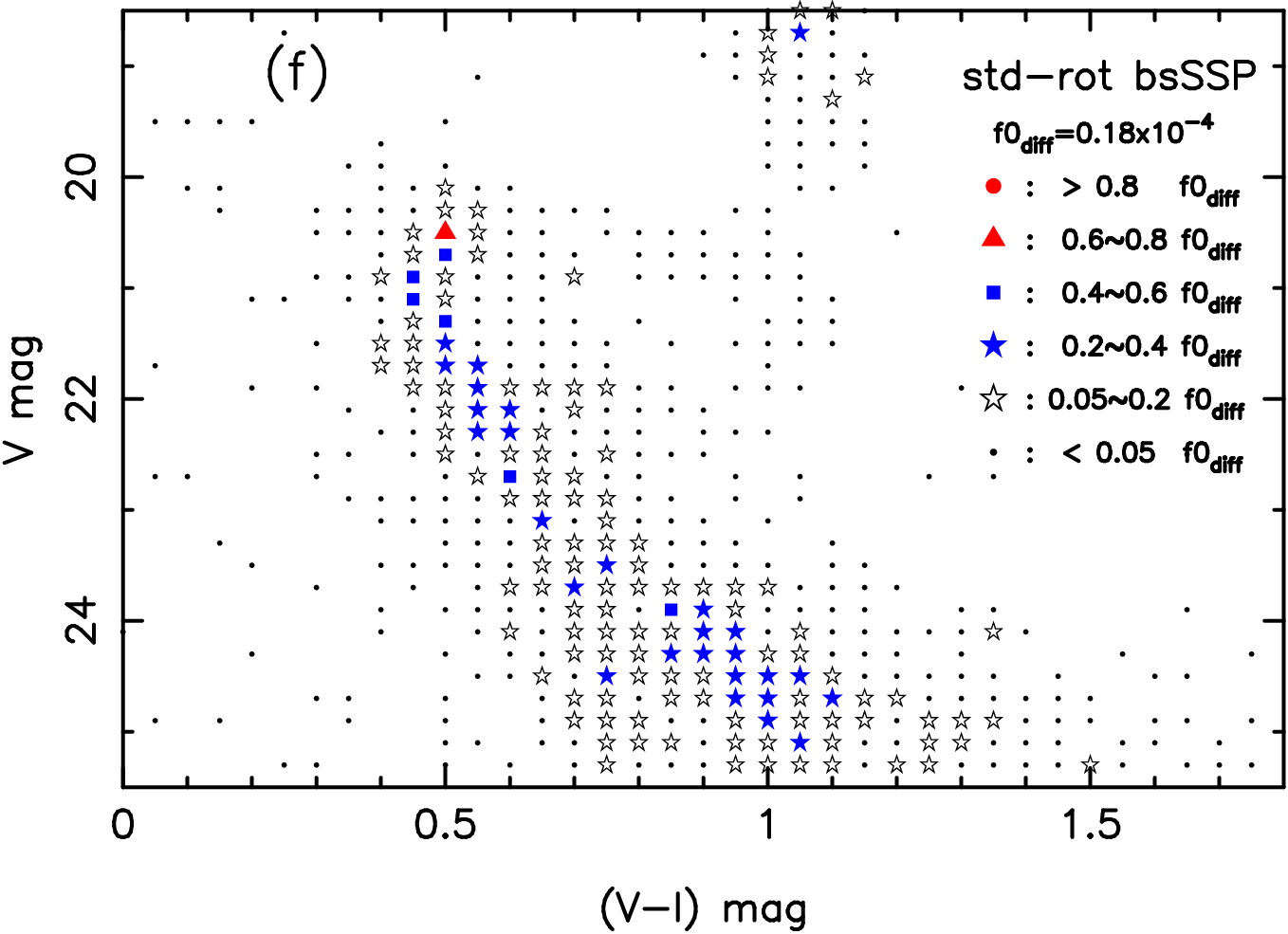}
\includegraphics[angle=-90,width=0.45\textwidth]{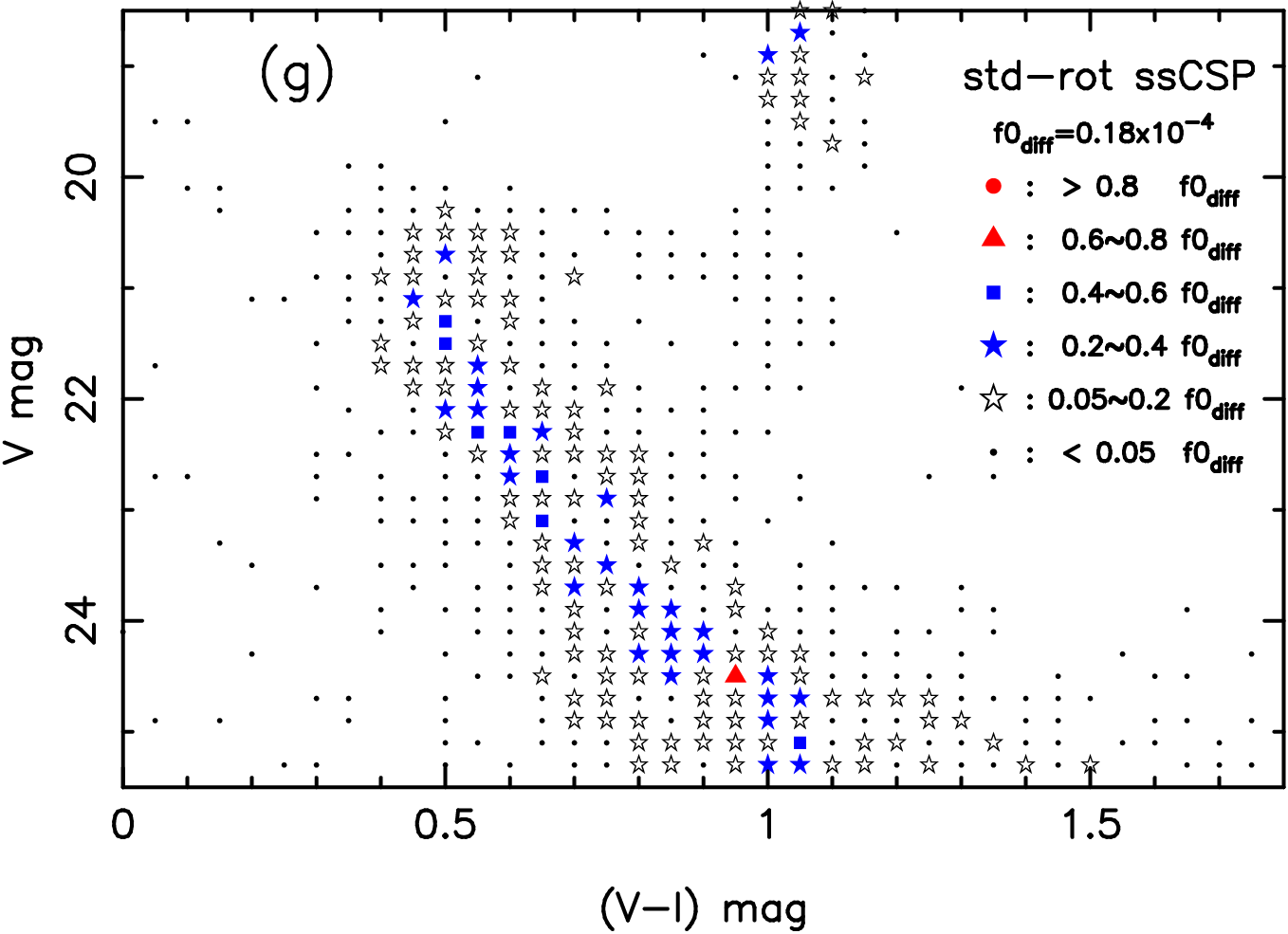}
\includegraphics[angle=-90,width=0.45\textwidth]{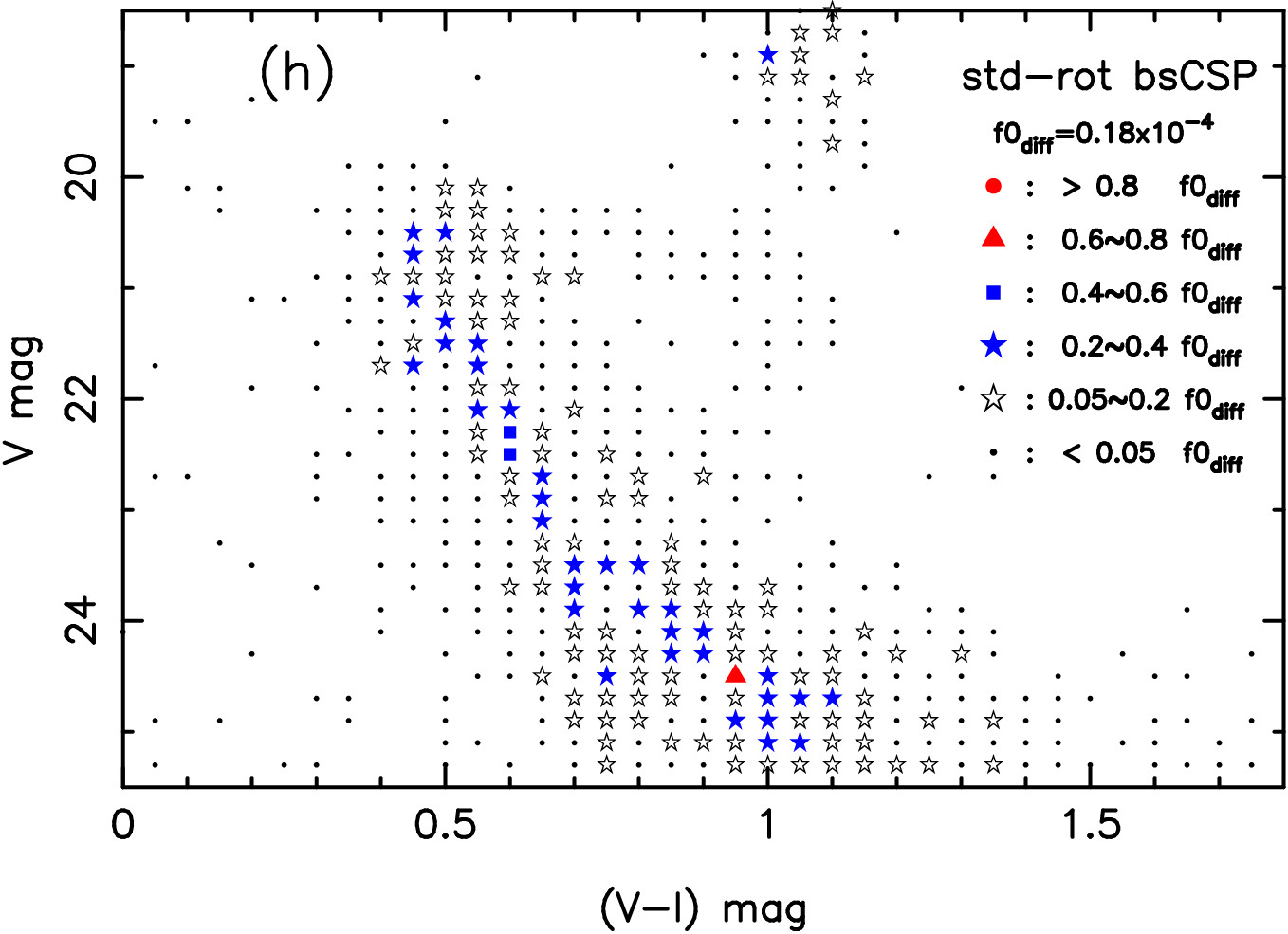}
\caption{\textbf{Star fraction difference between best-fit CMDs and observed CMD of NGC1651.
Best-fit CMDs are derived from different types of population models.
$f0_{\rm diff}$ is the maximum star fraction difference in all grids.}}
\end{figure}

 \begin{figure} %Fig 11.
\centering
\includegraphics[angle=-90,width=0.95\textwidth]{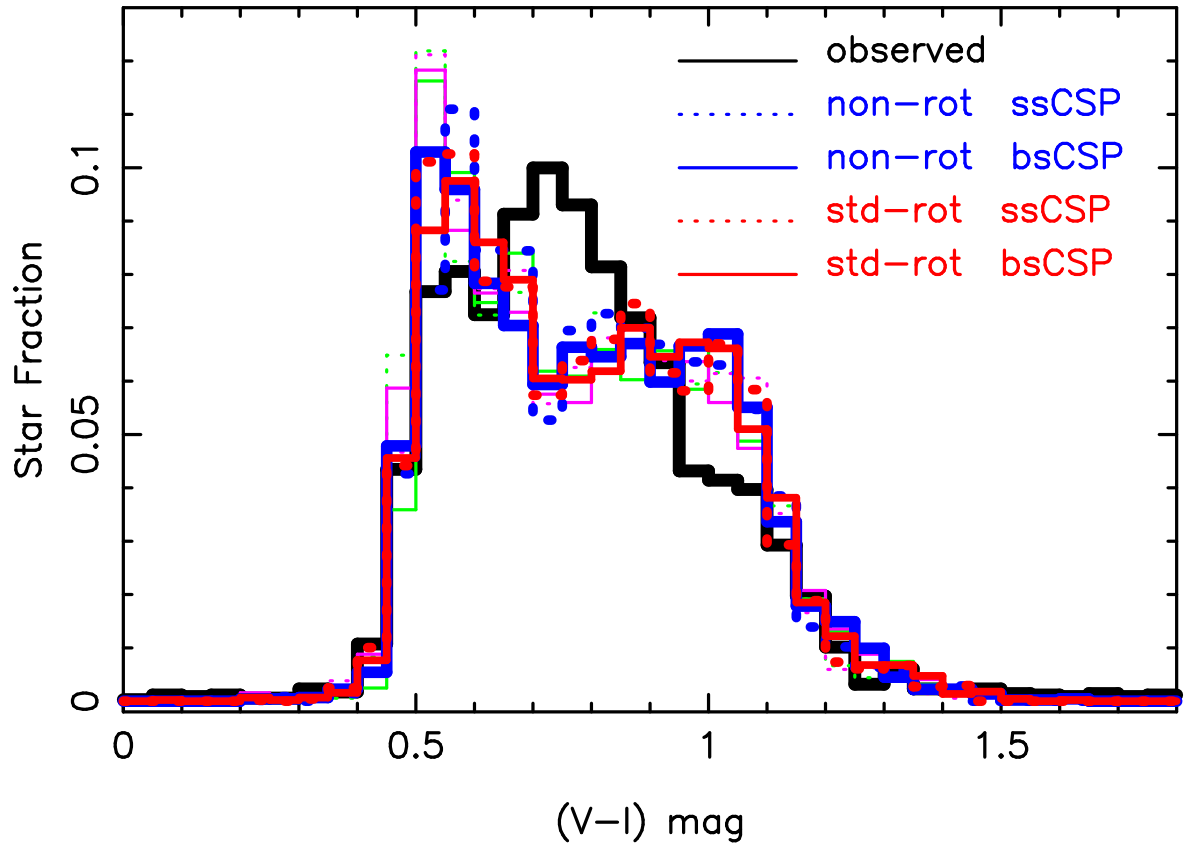}
\caption{Comparison of color distributions of observed and best-fit CMDs of NGC1651.}
\end{figure}

 \begin{figure} %Fig 12.
\centering
\includegraphics[angle=-90,width=0.95\textwidth]{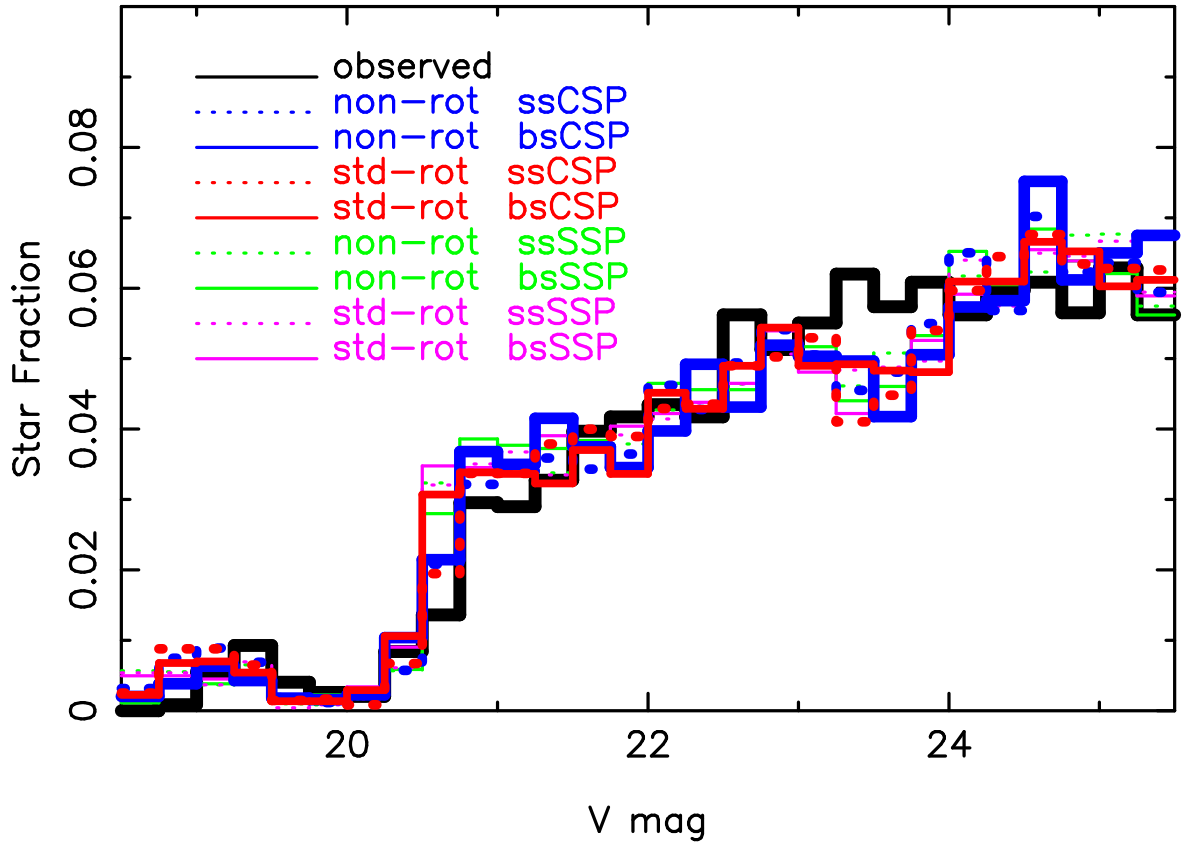}
\caption{Comparison of magnitude distributions of observed and best-fit CMDs of NGC1651.}
\end{figure}

\section{Conclusion and Discussion}

In this work, we used eight kinds of stellar population models to fit
the special CMD of LMC cluster NGC1651. Our results suggest that stellar binarity, rotation,
and age spread can demonstrably improve the goodness of CMD fit, but there is a degeneracy among their effects.
It is therefore difficult to judge the stellar population feature (e.g., age and star formation history) of this cluster using
traditional stellar population (TSP) models such as non-rotational single star simple stellar population (ssSSP)
and single star composite stellar population (ssCSP) models.
For future studies we strongly suggest using those advanced stellar population (ASP) models which take into account stellar binarity, rotation, and star formation history.

The result of this study shows that
cluster NGC1651 is of true distance modulus between 18.68 and 18.70\,mag,
$E(V-I)$ around 0.19\,mag, and binary fraction around 50
per cent.
We can conclude that both bsSSP and
bsCSP models can reproduce the main shape of CMD of NGC1651,
if the effect of stellar rotation is taken into account. A bsCSP
model with half binaries and 70 per cent rotational stars can fit the observed
CMD best among all the test models. The star formation history types (one or
a few star formations), stellar ages and star formation histories (or modes) from different kinds of
models are significantly different. Although our result prefers that NGC1651
is a bsCSP with rotational stars, it cannot exclude
the bsCSP model without rotators or bsSSP model that consists of rotational stars.
Further study (e.g., spectral study) is therefore needed to confirm the results. For instance, we can check how many and how fast the massive stars of a cluster are really rotating from the stellar spectra to judge the role of stellar rotation. Note that TSP models such as single star stellar population (ssSP) models,
seem significantly worse than binary star stellar populations
(bsSPs). Thus NGC1651 certainly contain some binaries,
and we need only to check how many rotators are included in a cluster like NGC1651.

In this work, we use the calculation of \cite{georgy2013} to
estimate the effects of stellar rotation. Although the value is not
accurate enough, the treatment is reasonable. First, in the work of \cite{georgy2013},
the variation in the angular momentum content is precisely tracked as it changes under the influence of stellar winds and mechanical mass loss.
The effects of initial rotation on the Hertzsprung-Russell diagram,
the evolution of surface rotation and abundances, and main sequence lifetime are computed simultaneously in this paper.
This makes their results more reliable than some other work (e.g., \citealt{bast09}).
Second, the observed CMD of NGC1651 is well reproduced when we adding the result of \cite{georgy2013}
into our binary star stellar population models.
Another important reason is that different works show the trend of
CMD change is correct when taking stellar rotation
into account (\citealt{bast09,Li12,yang2013}), although the observational work by \cite{platais12} and theoretical work by \cite{Girardi11} showed that rotation works in the opposite sense in a star cluster.
When we use the fitting formulae of \cite{bast09} to calculate rotational effects,
we also get the same conclusion.  Third, as pointed out above, the effects
of fraction of rotational stars, distribution of rotation rate and
fraction of binaries are degenerated in some cases.
Finally, although only the effects of massive rotational stars have been taken into account by rotational population
models because of the lack of evolutionary data of low-mass rotators, the treatment does not affect the final result too
much, as low-mass stars usually rotate much more slowly than massive ones.
Therefore, the treatment for the rotational effects of stars is reasonable at this moment,
and the main conclusion is accordingly reliable.

Furthermore, although \cite{Girardi13} argue that a spread of
rotation rate cannot lead to eMSTO, because rotation affects the
lifetime of stars, our work implies that the conclusion may strongly depend on stellar evolutionary models.
It shows that when taking the new result of \cite{georgy2013} for rotational effects,
the MS lifetime change not only reproduces our previous result from fitting formulae  \citep{Li12},
but also make the fit of std-rot bsCSP to eRC better.
Compared to our previous work,
the treatment in this work can reproduce the observed CMD better,
because most rotational stars leave MS later than non-rotating
ones. This changes the positions of components of binaries in a CMD when they leave MS and reach RC.
Because the evolutionary track and MS lifetime of a rotating binary depend on the masses, separation, and rotation rates of its component stars, rotation cannot lead to the same CMD position (see \citealt{Girardi11} for comparison) for a given age. The special CMD structure of a rotating-star population results from both the changes of evolutionary tracks and MS lifetime. Thereby, MS lifetime change also contribute to the final CMD, and rotation plays different roles in populations with and without binaries.
This is different from \citet{yang2013}, which claims that the increase in MS lifetime does not compensate for the changes in the evolutionary tracks.

Finally, although AST technique was used for estimating the
uncertainties in magnitudes of NGC1651,
more uncertainties exist and may affect the results.
In fact, certain other factors, including
uncertainties in metallicity, initial mass function, binary assumptions, calculation
of stellar evolution, atmosphere library, and contamination of field stars,
can also affect the CMDs of stellar populations.
Because the study of these uncertainties is not the purpose of this work, we did not
test their effects. However, further detailed studies on these uncertainties will be  very helpful. One can read many examples of this kind of detailed study, such as \cite{Dolphin2012, Dolphin2013}.

\acknowledgments  \textbf{The authors thank the referee for constructive comments,} Prof. Yan Li for suggesting the use of evolutionary tracks of rotational stars, Mr. Chengyuan Li for help with the use of
HSTphot, and Mr. Thomas Dodd for English checking. This work has been supported by the Chinese National
Science Foundation (Grant No. 11203005), Open Project of Key Laboratory for the Structure and Evolution of Celestial Objects,
Chinese Academy of Sciences (OP201304), and Foundation of Yunnan Education Department (No. 2010Z004).

%\bibliography{reftex}
%\bibliographystyle{apj}

\end{document}